# Effect of Marangoni stress on the bulk rheology of a dilute emulsion of surfactant-laden deformable droplets in linear flows

Shubhadeep Mandal, Sayan Das and Suman Chakraborty[a]

*Department of Mechanical Engineering, Indian Institute of Technology Kharagpur, West Bengal - 721302, India*

In the present study we analytically investigate the deformation and bulk rheology of a dilute emulsion of surfactant-laden droplets suspended in a linear flow. We use an asymptotic approach to predict the effect of surfactant distribution on the deformation of a single droplet as well as the effective shear and extensional viscosity for the dilute emulsion. The non-uniform distribution of surfactants due to the bulk flow results in the generation of a Marangoni stress which affects both the deformation as well as the bulk rheology of the suspension. The present analysis is done for the limiting case when the surfactant transport is dominated by the surface diffusion relative to surface convection. As an example, we have used two commonly encountered bulk flows, namely, uniaxial extensional flow and simple shear flow. With the assumption of negligible inertial forces present in either of the phases, we are able to show that both the surfactant concentration on the droplet surface as well as the ratio of viscosity of the droplet phase with respect to the suspending fluid has a significant effect on the droplet deformation as well as the bulk rheology. It is seen that increase in the non-uniformity in surfactant distribution on the droplet surface results in a higher droplet deformation and a higher effective viscosity for either of linear flows considered. For the case of simple shear flow, surfactant distribution is found to have no effect on the inclination angle, however, a higher viscosity ratio predicts the droplet to be more aligned towards the direction of flow

[a] E-mail address for correspondence: suman@mech.iitkgp.ernet.in

# I. INTRODUCTION

Emulsions, polymer blends and foams has been a leading area of research since a long time because of its wide application in foods, material processing and pharmaceuticals [1–3]. The droplets of the dispersed phase suspended in a carrier phase, are deformed, oriented and broken up during the processes that take place in the system. Hence the microstructure or the morphology of either of the phases changes depending on the shape, orientation, number density or distribution of the dispersed phase. The modified morphology helps us determine the various properties of the finished product, for example thermal, mechanical or chemical properties [1–4]. However to improve and control the emulsification and blending processes, analysis of a single droplet is necessary. The dynamics of suspended droplets also finds its wide application in different microfluidic devices [8–10]. Some of the common use of droplet emulsions in microfluidic devices can be found in cell encapsulation, reagent mixing, drug delivery and analytic detection [8,11–14].

The effect of surfactants on droplet dynamics have attracted researches since decades. For a surfactant-laden droplet the surface tension varies along the interface of the droplet due to a non-uniform distribution of surfactants along the droplet surface. A higher surfactant concentration results in low surface tension. Several experimental studies have shown that there is a close relationship between droplet deformation and local surfactant distribution along the droplet surface [15–20]. Vlahovska et al. [21], in their work, showed the effect of surfactant concentration on the deformation of the droplet suspended in a linear flow field. They considered two types of linear flows, namely, a uniaxial extensional flow and simple shear flow. They did their theoretical analysis for the limiting case when the surfactant transport on the droplet surface is dominated by convection rather than surface diffusion. Milliken et al. numerically investigated the effect of surfactant distribution on the migration, deformation and breakup of a droplet suspended in a uniaxial extensional flow [22]. For the case of an imposed linear flow the droplet elongates and there is a higher concentration of surfactants towards the two tips of the droplet. Thus a gradient in surface tension is created along the droplet surface with a lower surface tension near the tips of the droplet. This in turn gives rise to a Marangoni stress, which is main reason for the discontinuity in the tangential stress as well as the normal stress at the droplet interface. The rotational component of a simple shear flow, however, redistributes the surfactant thus decreasing the non-uniformity in surface tension and hence the deformation. The effect of rotation is more prominent for a higher viscosity of the droplet as compared to the suspending phase. Three dimensional numerical simulations have been done previously to investigate the dynamics of a surfactant laden droplet suspended in a simple shear flow [23–27]. Several numerical studies on Marangoni stress and shape deformation has been done previously for axisymmetric and two-dimensional flows [15,22,28–33]. In a recent study slender body theory has been used by Booty et al. to analytically study a highly deformable bubble [34].

A lot of analytical works on small deformation of the droplet can be found in the literature. However, most of them are done on a surfactant-free droplet [35–39]. For a droplet with a high viscosity as compared to the suspending fluid, deformations are usually small and hence an asymptotic approach is a better alternative than numerical simulation which are computationally expensive. Some of the noteworthy analytical works, done on the effect of surfactants on droplet deformation include the study done by Flummerfelt [40]. In this work they have derived a leading order perturbation theory where they have taken into account mass transfer from the bulk and also the effect of surface shear and dilatational viscosities. Stone and Leal also did a small deformation analysis where they included surface diffusion in their analysis [15]. A higher order solution is more of a challenge as all the boundary conditions needs to be evaluated at the deformed boundary.

A comprehensive analysis for the limiting case when the surfactant transport is dominated by surface diffusion is missing from the literature. This limiting case may arise in situations where the surface diffusivity of the surfactants is high. Experiments with a high value of surface diffusivity and a low imposed shear (or extensional) rate have been performed previously [27]. Towards this, we analytically study the effect of Marangoni stress on the dynamics of a surfactant-laden droplet in linear flows. The surfactants in the present problem are bulk-insoluble and get transported only along the droplet surface. As example we have considered two kind of linear flows: simple shear flow and uniaxial extensional flow. For each of the these linear flows we have obtained the deformed shape of the droplet and associated modification in the bulk rheology of a dilute emulsion.

## II. THEORETICAL MODEL

### A. Physical system

The physical system considered in this problem consists of a neutrally buoyant droplet suspended in a linear flow. The droplet has a radius of $a$ and is covered with bulk-insoluble surfactants, which are transported along the droplet interface due to surface diffusion and convection. The viscosity of the droplet as well as the suspending phase are $\mu_i$ and $\mu_e$, respectively. The subscript '$i$' is used to denote the droplet phase quantities, whereas the subscript '$e$' refers to the quantities related to the suspending phase. In the present study we have considered the droplet to be suspended in a linear flow, which may be uniaxial extensional flow or simple shear flow. This imposed flow field $\bar{\mathbf{u}}_\infty$ can be represented mathematically in a general form as

$$\bar{\mathbf{u}}_\infty = \left(\bar{\mathbf{D}}_\infty + \bar{\mathbf{\Omega}}_\infty\right) \cdot \bar{\mathbf{x}} = \bar{\mathbf{D}}_\infty \cdot \bar{\mathbf{x}} + \frac{1}{2}\left(\bar{\mathbf{\omega}}_\infty \times \bar{\mathbf{x}}\right), \tag{1}$$

where $\bar{\mathbf{D}}_\infty$ is the rate of strain tensor, $\bar{\mathbf{x}}$ is the position vector, $\bar{\mathbf{\Omega}}_\infty$ is the vorticity tensor and $\bar{\mathbf{\omega}}_\infty$ is the vorticity vector. For a simple shear flow we have in the above expression

$$\bar{\mathbf{D}}_\infty = \frac{\dot\gamma}{2}\begin{bmatrix} 0 & 1 & 0 \\ 1 & 0 & 0 \\ 0 & 0 & 0 \end{bmatrix},\ \bar{\mathbf{\Omega}}_\infty = \frac{\dot\gamma}{2}\begin{bmatrix} 0 & 1 & 0 \\ -1 & 0 & 0 \\ 0 & 0 & 0 \end{bmatrix}, \tag{2}$$

where $\dot\gamma$ is the shear rate. For a uniaxial extensional flow on the other hand we have

$$\bar{\mathbf{D}}_\infty = \frac{\dot\gamma}{2}\begin{bmatrix} -1 & 0 & 0 \\ 0 & -1 & 0 \\ 0 & 0 & 2 \end{bmatrix},\ \bar{\mathbf{\Omega}}_\infty = \mathbf{0}. \tag{3}$$

A schematic of the system is given in Fig. 1, where we have only shown the case in which a surfactant-laden droplet is suspended in a simple shear flow. A spherical coordinate system $(\bar r,\theta,\varphi)$ and a cartesian coordinate system $(\bar x,\bar y,\bar z)$ is attached to the centroid of the droplet. In the absence of any surfactant, that is for a clean droplet, the surface tension of the suspended droplet is constant and is denoted by $\bar\sigma_c$. On the other hand, a surfactant-laden droplet suspended in a quiescent fluid with no imposed flow has a uniform surfactant distribution $(\bar\Gamma_{eq})$ with a corresponding constant surface tension, $\bar\sigma_{eq}$. Presence of an imposed flow, however, renders the surfactant distribution non-uniform which results in the variation of surface tension along the droplet surface. This variation in surface tension is responsible for the generation of Marangoni stress, which not only causes deformation of the droplet but also drives fluid flow.

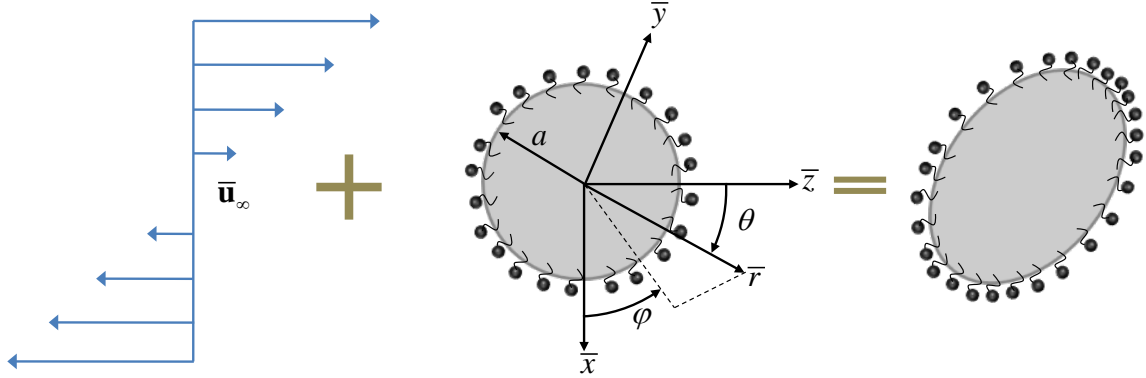

Fig. 1. Schematic of a surfactant-laden droplet of radius $a$ suspended in a linear flow. As an example we have shown the background flow to be a simple shear flow. Both the spherical $(\bar r,\theta,\varphi)$ as well as the Cartesian coordinates $(\bar x,\bar y,\bar z)$ are shown. Either of them are fixed to the centroid of the droplet.

The aim of the present study is to analyze the effect of surfactants on the droplet deformation for both types of imposed flows: simple shear flow and uniaxial extensional flow. We also, in either of these cases, investigate the effect of surfactant concentration on suspension rheology. Towards this, we find out the effective shear viscosity and effective extensional viscosity of a dilute suspension of droplets in a simple shear flow and uniaxial extensional flow, respectively.

## B. Important assumptions

In order to analytically solve the above problem, some assumptions have to be made in order to simplify the governing equations as well as boundary conditions for flow field. These assumptions are as follows:

(i) The pressure, viscous as well as surface tension forces acting on the droplet are assumed to be more dominating in comparison to the inertia force. In other words, the flow Reynolds number, $Re = \rho \dot{\gamma} a^2 / \mu_e$, is assumed to be small ($Re \ll 1$). Here $\rho$ is the density of either of the phases.

(ii) The transport of surfactants is assumed to take place only along the droplet surface, without any net flux into either of the phases. That is, the surfactant is considered to be bulk-insoluble.

(iii) The surface tension at the droplet interface is assumed to be linearly related to the local surfactant concentration.

(iv) The droplet dynamics is assumed to be unaffected by any bounding walls, if present .That is, the droplet is assumed to be much smaller as compared to the distance between the walls.

(v) The droplet is assumed to be approximately spherical, that is, only small deformations are considered. For creeping flow, the droplet deformation is governed by the magnitude of the capillary number $\left(Ca^* = \mu_e \dot{\gamma} a / \bar{\sigma}_c\right)$, which is the ratio of viscous force to the surface tension force acting on the droplet. In this problem we assume small deformation of the droplet only, which restricts us to small values of capillary number ($Ca^* \ll 1$).

## C. Dimensional form of governing equations and boundary conditions

We first start by stating the governing equations for flow field. The flow field, under the above assumptions is governed by the Stokes and continuity equations. The dimensional form of these equations can be written as

$$\left.\begin{array}{l}-\bar{\nabla}\bar{p}_i + \mu_i \bar{\nabla}^2 \bar{\mathbf{u}}_i = \mathbf{0}, \ \bar{\nabla}\cdot\bar{\mathbf{u}}_i = 0, \\ -\bar{\nabla}\bar{p}_e + \mu_e \bar{\nabla}^2 \bar{\mathbf{u}}_e = \mathbf{0}, \ \bar{\nabla}\cdot\bar{\mathbf{u}}_e = 0,\end{array}\right\} \quad (4)$$

where $\bar{\mathbf{u}}_{i,e}$ is the velocity field and $\bar{p}_{i,e}$ is the pressure field. The 'overbar' in the above equation is used to denote dimensional quantities, while the dimensionless quantities and material properties are denoted without overbar.

The far-field condition satisfied by the velocity as well as pressure field outside the droplet $\left(\bar{\mathbf{u}}_e, \bar{p}_e\right)$ is given by

$$\text{as } \bar{r} \to \infty, \quad \bar{\mathbf{u}}_e = \bar{\mathbf{u}}_\infty,$$
$$\text{as } \bar{r} \to \infty, \quad \bar{p}_e = \bar{p}_\infty, \tag{5}$$

where $\bar{\mathbf{u}}_\infty$ and $\bar{p}_\infty$ is the velocity and pressure at far-field, respectively. The expression for $\bar{\mathbf{u}}_\infty$ is provided in Eq. (1). Inside the droplet, the velocity and pressure fields $(\bar{\mathbf{u}}_i, \bar{p}_i)$ are bounded at the centroid of the droplet, $\bar{r} = 0$. The boundary conditions at the interface of the droplet $(\bar{r} = \bar{r}_s)$, where $\bar{r}_s$ is the dimensional radial position of the droplet interface, consists of the no-slip condition, the kinematic boundary condition and finally the balance between hydrodynamic and Marangoni stresses. The dimensional form of the interfacial boundary conditions are of the form

$$\left.\begin{array}{l} \text{at } \bar{r} = \bar{r}_s, \quad \bar{\mathbf{u}}_i = \bar{\mathbf{u}}_e, \\ \text{at } \bar{r} = \bar{r}_s, \quad \bar{\mathbf{u}}_i \cdot \mathbf{n} = \bar{\mathbf{u}}_e \cdot \mathbf{n} = 0, \\ \text{at } \bar{r} = \bar{r}_s, \quad (\bar{\boldsymbol{\tau}}_e \cdot \mathbf{n} - \bar{\boldsymbol{\tau}}_i \cdot \mathbf{n}) = -\bar{\nabla}_s \bar{\sigma} + \bar{\sigma}(\bar{\nabla} \cdot \mathbf{n})\mathbf{n}, \end{array}\right\} \tag{6}$$

where $\bar{\nabla}_s = (\mathbf{I} - \mathbf{nn}) \cdot \bar{\nabla}$ is the surface gradient tensor and $\mathbf{n}$ is the unit vector normal to the droplet surface and is given by

$$\mathbf{n} = \frac{\bar{\nabla} \bar{F}}{|\bar{\nabla} \bar{F}|}, \tag{7}$$

where $\bar{F} = \bar{r} - \bar{r}_s$ is the equation for the droplet surface. and $\bar{\boldsymbol{\tau}}_e$ and $\bar{\boldsymbol{\tau}}_i$ are the external and internal viscous/hydrodynamic stress tensors, given by

$$\left.\begin{array}{l} \bar{\boldsymbol{\tau}}_i = -\bar{p}_i \mathbf{I} + \mu_i \left[ \bar{\nabla} \bar{\mathbf{u}}_i + (\bar{\nabla} \bar{\mathbf{u}}_i)^T \right], \\ \bar{\boldsymbol{\tau}}_e = -\bar{p}_e \mathbf{I} + \mu_e \left[ \bar{\nabla} \bar{\mathbf{u}}_e + (\bar{\nabla} \bar{\mathbf{u}}_e)^T \right]. \end{array}\right\} \tag{8}$$

Under the assumption of bulk insolubility, the surfactant transport at steady state is governed by a convection-diffusion equation which can be written as [41]

$$\bar{\nabla}_s \cdot (\bar{\mathbf{u}}_s \bar{\Gamma}) = D_s \bar{\nabla}_s^2 \bar{\Gamma}, \tag{9}$$

where $\bar{\Gamma}$ is dimensional local surfactant concentration, $\bar{\mathbf{u}}_s$ is the interfacial velocity and $D_s$ is the surface diffusivity of the surfactants. The above surfactant transport equation is coupled with the flow field. This clearly suggests that the local surfactant concentration has to solved in conjunction with the flow field. Finally the dimensional equation of state correlating the surface tension with the local surfactant concentration is given by [21]

$$\bar{\sigma} = \bar{\sigma}_c - R_g \bar{T}_o \bar{\Gamma}, \tag{10}$$

where $\bar{T}_o$ is any reference temperature and $R_g$ is the universal gas constant.

### D. Dimensionless form of governing equations and boundary conditions

We now move forward in deriving the dimensionless form of the governing differential equations and relevant boundary conditions. Towards this, we use the following characteristic scales

$$\left. \begin{array}{l} r = \bar{r}/a,\ \mathbf{u} = \bar{\mathbf{u}}/\dot{\gamma}a,\ \Gamma = \bar{\Gamma}/\bar{\Gamma}_{eq},\ \sigma = \bar{\sigma}/\bar{\sigma}_c, \\ p = \bar{p}/(\mu_e \dot{\gamma}),\ \boldsymbol{\tau} = \bar{\boldsymbol{\tau}}/(\mu_e \dot{\gamma}) \end{array} \right\} \tag{11}$$

Different entities which will be useful while deriving the dimensionless form of governing equations and boundary conditions are (i) the viscosity ratio, $\lambda = \mu_i/\mu_e$, (ii) the surface Péclet number, $Pe_s = \dot{\gamma}a^2/D_s$, which signifies the relative importance of convection in the transport of surfactants along the droplet surface with respect to surface diffusion, (iii) the elasticity number, $\beta = \bar{\Gamma}_{ref} R\bar{T}_o/\bar{\sigma}_c = -d(\bar{\sigma}/\bar{\sigma}_c)/d\bar{\Gamma}$, which indicates the sensitivity of surface tension towards local surfactant concentration, and (iv) the modified capillary number, $Ca = Ca^*/(1-\beta)$. The main reason for the use of the modified capillary number is that it is defined based on the equilibrium surface tension for a surfactant-laden droplet $\left(\bar{\sigma}_{eq} = \bar{\sigma}_c(1-\beta)\right)$ rather than the surface tension for a clean droplet $(\bar{\sigma}_c)$. Such a choice adds to our convenience in further calculation.

Thus using the above non-dimensional scheme, the following set of non-dimensional governing differential equations for flow field are obtained

$$\left. \begin{array}{l} -\nabla p_i + \lambda \nabla^2 \mathbf{u}_i = \mathbf{0},\ \nabla \cdot \mathbf{u}_i = 0, \\ -\nabla p_e + \nabla^2 \mathbf{u}_e = \mathbf{0},\ \nabla \cdot \mathbf{u}_e = 0, \end{array} \right\} \tag{12}$$

and the relevant boundary conditions are given by

$$\left.\begin{aligned}
&\text{at } r \to \infty, \ (\mathbf{u}_e, p_e) = (\mathbf{u}_\infty, p_\infty), \\
&\mathbf{u}_i \text{ is bounded at } r = 0, \\
&\text{at } r = r_s, \ \mathbf{u}_i \cdot \mathbf{n} = \mathbf{u}_e \cdot \mathbf{n} = 0, \\
&\text{at } r = r_s, \ \mathbf{u}_i = \mathbf{u}_e, \\
&\text{at } r = r_s, \ (\boldsymbol{\tau}_e \cdot \mathbf{n} - \boldsymbol{\tau}_i \cdot \mathbf{n}) = \frac{\beta}{(1-\beta)Ca} \nabla_s \Gamma + \frac{\sigma}{Ca}(\nabla \cdot \mathbf{n}).
\end{aligned}\right\} \quad (13)$$

The last of the above set boundary conditions are obtained as a result of substitution of the non-dimensional form of equation of state, given by

$$\sigma = 1 - \beta \Gamma. \quad (14)$$

The surface tension based on the modified capillary number can be written in the following form

$$\sigma = \frac{\bar{\sigma}}{\bar{\sigma}_c(1-\beta)}. \quad (15)$$

From the above equation it can be said that $0 < \beta < 1$. The dimensionless surfactant transport equation is given by

$$Pe_s \nabla_s \cdot (\mathbf{u}_s \Gamma) = \nabla_s^2 \Gamma. \quad (16)$$

The mass conservation constraint to be satisfied by the local surfactant concentration along the droplet surface can expressed in the following form

$$\int_{\varphi=0}^{2\pi} \int_{\theta=0}^{\pi} \Gamma(\theta, \varphi) \sin\theta \, d\theta \, d\varphi = 4\pi. \quad (17)$$

It is clear from the above set of governing equations (12) and (16), that an exact solution for the flow field as well as surfactant concentration is not possible analytically due to the coupled nature of the flow and surfactant transport. In addition to this, the unknown deformed shape of the droplet renders the set of governing equations and boundary conditions non-linear in nature. Fortunately, an asymptotic approach just serves the purpose. [42,43] We use a small deformation theory with *Ca* as perturbation parameter and express the deformed shape of the droplet as $F = r - r_s$. The solution is thus obtained for the limiting case of low surface Péclet number, $Pe_s \ll 1$.

## III. ASYMPTOTIC SOLUTION

In the limiting case of $Pe_s \ll 1$, the surfactant transport is dominated by the surface diffusion in comparison to convection at the droplet interface. The magnitude of $Pe_s$ is taken to be of the same order as that of capillary number $(Ca)$, that is $Pe_s \sim Ca$. This can be written in the following form

$$Pe_s = kCa, \tag{18}$$

where $k = a\bar{\sigma}_c(1-\beta)/\mu_e D_s$ is called the property parameter as it depends on the various material properties. It has a magnitude of $O(1)$. Thus the droplet deformation is solely a function of $Ca$ for any given values of $k$ and $\beta$.

We thus choose the capillary number as the perturbation parameter. All flow variables can thus be expanded in a power series as follows

$$\psi = \psi^{(0)} + \psi^{(Ca)} Ca + O(Ca^2), \tag{19}$$

where $\psi$ is any generic flow variable. The first term in this expansion represents the leading order term that is the flow variable, $\psi$ for no deformation. All the other terms in this expansion are $O(Ca)$ or higher order correction terms due to deformation of the droplet. The surfactant concentration on the other hand is expanded as follows

$$\Gamma = 1 + \Gamma^{(0)} Ca + \Gamma^{(Ca)} Ca^2 + O(Ca^3). \tag{20}$$

The surfactant concentration obtained at each order of perturbation should always satisfy the mass conservation constraint on the droplet surface as given in Eq. (17).

Towards obtaining an asymptotic solution we express all the different quantities involved in terms of spherical harmonics. The local surfactant concentration is expressed in the form of an infinite series in terms of spherical surface harmonics as follows

$$\Gamma = \sum_{n=0}^{\infty} \sum_{m=0}^{n} \left[ \Gamma_{n,m} \cos(m\varphi) + \hat{\Gamma}_{n,m} \sin(m\varphi) \right] P_{n,m}(\cos\theta), \tag{21}$$

where $P_{n,m}(\cos\theta)$ are the associated Legendre polynomials of order $m$ and degree $n$. The unknown coefficients, $\Gamma_{n,m}$ and $\hat{\Gamma}_{n,m}$ are found out by solving the surfactant transport equation.

As both the flow inside as well as outside the droplet satisfies the Stokes equation, the use of general Lamb's solution can be made to find out the velocity and pressure field in either of the phases. The general Lamb's solution for velocity and pressure field inside the droplet in terms of growing spherical solid harmonics is given by

$$\mathbf{u}_i = \sum_{n=1}^{\infty} \left[ \nabla \times (\mathbf{r}\chi_n) + \nabla \Phi_n + \frac{n+3}{2(n+1)(2n+3)\lambda} r^2 \nabla p_n - \frac{n}{(n+1)(2n+3)\lambda} \mathbf{r} p_n \right], \quad (22)$$

$$p_i = \sum_{n=0}^{\infty} p_n,$$

where $p_n, \chi_n$ and $\Phi_n$ are growing spherical harmonics, the expressions of which can be found in the work done by Haber and Hetsroni [44]. In a similar manner, the velocity and pressure fields outside the droplet can be expressed in terms of decaying solid harmonics as

$$\mathbf{u}_e = \mathbf{u}_\infty + \sum_{n=1}^{\infty} \left[ \nabla \times (\mathbf{r}\chi_{-n-1}) + \nabla \Phi_{-n-1} - \frac{n-2}{2n(2n-1)} r^2 \nabla p_{-n-1} + \frac{n+1}{n(2n-1)} \mathbf{r} p_{-n-1} \right], \quad (23)$$

$$p_e = p_\infty + \sum_{n=0}^{\infty} p_{-n-1},$$

where $p_{-n-1}, \chi_{-n-1}$ and $\Phi_{-n-1}$ decaying solid spherical harmonics, expression of which can also be found in the work of Haber and Hetsroni [44].

The velocity and pressure fields can thus be calculated with the help of the boundary conditions at the droplet interface namely the kinematic boundary condition, the no-slip condition and the tangential stress balance. The tangential stress boundary condition can be obtained from the stress balance condition as given in the last of the equations in (13). The tangential stress boundary condition signifies the jump in tangential stress at the droplet interface due to the presence of the surfactant-induced Marangoni stress. This boundary condition can be written in the following form

$$\text{at } r = r_s, \quad (\boldsymbol{\tau}_e \cdot \mathbf{n} - \boldsymbol{\tau}_i \cdot \mathbf{n}) \cdot (\mathbf{I} - \mathbf{nn}) = \frac{\beta}{(1-\beta)Ca} (\nabla_s \Gamma) \cdot (\mathbf{I} - \mathbf{nn}) \quad (24)$$

where $\mathbf{I}$ is an identity tensor and $r_s$ represents the radial distance of the deformed surface of the drop and can be written in the following form

$$r_s = 1 + Ca\, g^{(Ca)} + Ca^2 g^{(Ca^2)} + O(Ca^3), \quad (25)$$

where $g^{(Ca)}$ and $g^{(Ca^2)}$ are $O(Ca)$ and $O(Ca^2)$ correction to the spherical shape of the droplet.

We next proceed towards obtaining the solution for flow field with the help of the following steps

(i) We first substitute expressions (18), (19) and (20) into equations (12), (13) and (16) and thus obtain the governing differential equations and boundary conditions for leading-order and $O(Ca)$ perturbation.

(ii) The flow field boundary conditions (other than the normal stress boundary condition) and the surfactant transport equation for leading-order are next solved simultaneously to calculate the spherical harmonics. We thus get the leading order solution for surfactant concentration.

(iii) We next substitute the expressions for solid spherical harmonics in equations (22) and (23), to obtain the pressure and velocity fields both outside as well as inside the droplet.

(iv) With the leading order solution at hand, we further calculate the $O(Ca)$ deformation which can be obtained from the normal stress balance at the deformed interface of the droplet, $r = r_s$. The normal stress balance obtained from the stress balance equation given in equation (13) is written as

$$\text{at } r = r_s, \ (\boldsymbol{\tau}_e \cdot \mathbf{n} - \boldsymbol{\tau}_i \cdot \mathbf{n}) \cdot \mathbf{n} = \frac{\sigma}{Ca}(\nabla \cdot \mathbf{n}) \tag{26}$$

(v) Thus expanding the stress tensors as shown in Eq. (19), substituting the expression of $\mathbf{n}$ from Eq. (7) and $\sigma$ from Eq. (14) and applying the orthogonality condition for the associate Legendre polynomials on either sides of the normal stress balance we get the expression for $O(Ca)$ correction to the droplet shape which is given by

$$g^{(Ca)} = \sum_{n=0}^{\infty} \sum_{m=0}^{n} \left[ L_{n,m}^{(Ca)} \cos(m\varphi) + \hat{L}_{n,m}^{(Ca)} \sin(m\varphi) \right] P_{n,m}(\cos\theta), \tag{27}$$

where $L_{n,m}^{(Ca)}$ and $\hat{L}_{n,m}^{(Ca)}$ are constant coefficients.

(vi) With the leading order solution as well as $O(Ca)$ deformation at our disposal, we proceed further towards calculating the $O(Ca)$ solution for flow field and $O(Ca^2)$ correction to the droplet shape. We first start by deriving the $O(Ca)$ boundary conditions and surfactant transport equation at the deformed surface $(r = r_s)$ of the droplet.

(vii) These equations are then solved simultaneously to obtain the $O(Ca)$ surfactant concentration as well as the flow field.

(viii) Next we again use the orthogonality condition for associated Legendre polynomials on the either sides of the $O(Ca)$ normal stress balance to calculate the $O(Ca^2)$ correction to the droplet shape. This correction in droplet shape is given by

$$g^{(Ca^2)} = \sum_{n=0}^{\infty} \sum_{m=0}^{n} \left[ L_{n,m}^{(Ca^2)} \cos(m\varphi) + \hat{L}_{n,m}^{(Ca^2)} \sin(m\varphi) \right] P_{n,m}(\cos\theta), \tag{28}$$

We now put forward the expressions for the surfactant concentration as well as $O(Ca)$ and $O(Ca^2)$ correction to the droplet shape for the following two separate cases: (a) imposed uniaxial extensional flow and (b) an imposed simple shear flow.

## A. Uniaxial extensional flow field

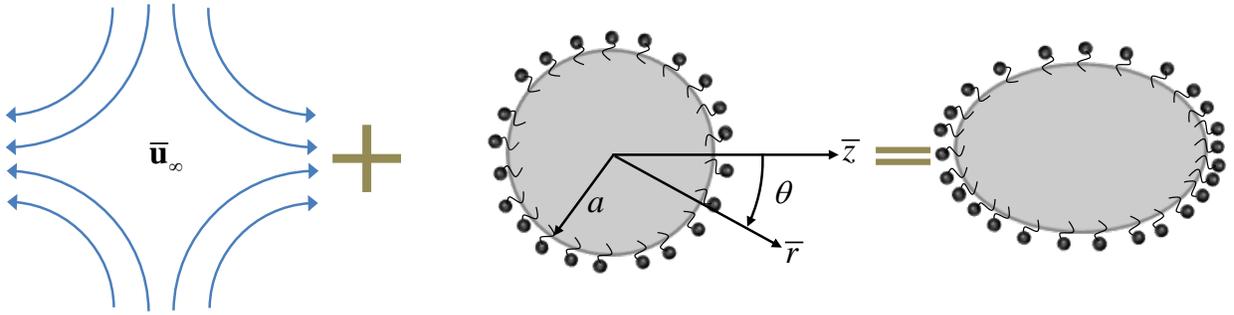

Fig. 2. Schematic of a surfactant-laden droplet suspended in a uniaxial extensional flow. Both the imposed flow field as well as non-uniform distribution of surfactants are responsible for the deformation of the droplet.

A schematic for the case of a surfactant-laden droplet suspended in an uniaxial extensional flow field is shown below in Fig. 2. The velocity as well as the pressure fields for this special case are given in Appendix A. The surfactant concentration at the droplet interface when a uniaxial extensional flow is imposed in the far-field is given by Eq. (20). The expressions for $\Gamma^{(0)}$ and $\Gamma^{(Ca)}$ are of the form

$$\left.\begin{array}{l}\Gamma^{(0)} = \Gamma_{2,0}^{(0)} P_{2,0}, \\ \text{and, } \Gamma^{(Ca)} = \Gamma_{2,0}^{(Ca)} P_{2,0} + \Gamma_{4,0}^{(Ca)} P_{4,0}\end{array}\right\} \tag{29}$$

where the above constant coefficients are obtained as

$$\left.\begin{aligned}
\Gamma_{2,0}^{(0)} &= \frac{5k}{2}\left\{\frac{1-\beta}{5+k\beta+5\lambda-5\beta-5\lambda\beta}\right\}, \\
\Gamma_{2,0}^{(Ca)} &= -25k\frac{\left\{g1_{2,0}\beta^3 + g2_{2,0}\beta^2 + g3_{2,0}\beta + g4_{2,0}\right\}}{112(\beta k - 5\beta - 5\beta\lambda + 5 + 5\lambda)^3}, \\
\Gamma_{4,0}^{(Ca)} &= -45k\frac{\left\{g1_{4,0}\beta^3 + g2_{4,0}\beta^2 + g3_{4,0}\beta + g4_{4,0}\right\}}{112(\beta k - 5\beta - 5\beta\lambda + 5 + 5\lambda)^2(\beta k - 9\beta - 9\beta\lambda + 9\lambda + 9)},
\end{aligned}\right\} \quad (30)$$

The constants present in the above equation are provided in Appendix A.

The deformed shape of the droplet when it is suspended in a uniaxial extensional flow field is given below

$$r_s = 1 + Ca\left\{L_{2,0}^{(Ca)}P_{2,0}\right\} + Ca^2\left\{L_{0,0}^{(Ca^2)} + L_{2,0}^{(Ca^2)}P_{2,0} + L_{4,0}^{(Ca^2)}P_{4,0}\right\}, \quad (31)$$

where $L_{0,0}^{(Ca^2)}$ is included in the above $O(Ca^2)$ correction to take into consideration the volume conservation constraint. The volume conservation constraint is given by

$$\int_{\varphi=0}^{2\pi}\int_{\theta=0}^{\pi}\int_{r=0}^{r_s} r^2 dr d\theta d\varphi = \frac{4\pi}{3}. \quad (32)$$

Thus using the above volume conservation condition, $L_{0,0}^{(Ca^2)}$ is found out to be

$$L_{0,0}^{(Ca^2)} = -\frac{1}{5}L_{2,0}^{(Ca)} \quad (33)$$

The constant coefficients present in Eq. (31) are given below

$$L_{2,0}^{(Ca)} = \frac{5}{8}\left(\frac{16-16\beta+19\lambda+4k\beta-19\lambda\beta}{5+k\beta+5\lambda-5\beta-5\lambda\beta}\right), \quad (34)$$

$$L_{2,0}^{(Ca^2)} = \frac{\left\{\begin{array}{l}\left(a_0^{(20)}\lambda^3 + a_1^{(20)}\lambda^2 + a_2^{(20)}\lambda + a_3^{(20)}\right)\beta^3 + \left(b_0^{(20)}\lambda^3 + b_1^{(20)}\lambda^2 + b_2^{(20)}\lambda + b_3^{(20)}\right)\beta^2 \\ +\left(c_0^{(20)}\lambda^3 + c_1^{(20)}\lambda^2 + c_2^{(20)}\lambda + c_3^{(20)}\right)\beta + 25(19\lambda+16)(601\lambda^2 + 893\lambda + 256)\end{array}\right\}}{448(\beta k - 5\beta - 5\beta\lambda + 5 + 5\lambda)^3}, \quad (35)$$

$$L_{4,0}^{(Ca^2)} = \frac{\begin{cases}\left(a_0^{(40)}\lambda^3 + a_1^{(40)}\lambda^2 + a_2^{(40)}\lambda + a_3^{(40)}\right)\beta^3 + \left(b_0^{(40)}\lambda^3 + b_1^{(40)}\lambda^2 + b_2^{(40)}\lambda + b_3^{(40)}\right)\beta^2 \\ + \left(c_0^{(40)}\lambda^3 + c_1^{(40)}\lambda^2 + c_2^{(40)}\lambda + c_3^{(40)}\right)\beta + 15(751\lambda + 656)(19\lambda + 16)(\lambda + 1)\end{cases}}{224(\beta k - 5\beta - 5\beta\lambda + 5 + 5\lambda)^2(\beta k - 9\beta - 9\beta\lambda + 9\lambda + 9)}, \quad (36)$$

where the expressions for the constants $a_0^{(20)} - a_3^{(20)}$, $b_0^{(20)} - b_3^{(20)}$, $c_0^{(20)} - c_3^{(20)}$, $a_0^{(40)} - a_3^{(40)}$, $b_0^{(40)} - b_3^{(40)}$ and $c_0^{(40)} - c_3^{(40)}$ are given in Appendix A. All the other coefficients $L_{n,m}^{(Ca)}, L_{n,m}^{(Ca^2)}$ are zero. Although the present study is for the case of $k \sim O(1)$, we still obtain the results of Vlahovska et al. [21] on substitution of $k \to \infty$. The deformation can be conveniently quantified for the case of small deformation $(Ca \ll 1)$ with the help of a deformation parameter, $D_{fe}$ which, for the case of an extensional flow field is given by

$$D_{fe} = \frac{r_s(\theta = 0) - r_s(\theta = \pi/2)}{r_s(\theta = 0) + r_s(\theta = \pi/2)}. \quad (37)$$

### B. Simple shear flow field

The schematic for the special case when the imposed flow field is a simple shear flow is shown in Fig. 1. The expressions for the velocity and pressure field are given in Appendix B. The constant coefficients in the expression of surfactant concentration of a droplet in an simple shear flow field, as shown in Eq. (20), is given by

$$\left.\begin{aligned}\Gamma^{(0)} &= \hat{\Gamma}_{2,2}^{(0)}\sin(2\varphi)P_{2,2}, \text{ and} \\ \Gamma^{(Ca)} &= \Gamma_{2,0}^{(Ca)}P_{2,0} + \Gamma_{4,0}^{(Ca)}P_{4,0} + \Gamma_{2,2}^{(Ca)}\cos(2\varphi)P_{2,2} + \Gamma_{4,4}^{(Ca)}\cos(4\varphi)P_{4,4},\end{aligned}\right\} \quad (38)$$

where the constant coefficients present in the above equation is given in Appendix B.

The deformed shape of the droplet is next given by

$$r_s = \begin{bmatrix}1 + Ca\left\{\hat{L}_{2,2}^{(Ca)}\sin(2\varphi)P_{2,2}\right\} \\ + Ca^2\left\{L_{0,0}^{(Ca^2)} + L_{2,0}^{(Ca^2)}P_{2,0} + L_{4,0}^{(Ca^2)}P_{4,0} + L_{2,2}^{(Ca^2)}\cos(2\varphi)P_{2,2} + L_{4,4}^{(Ca^2)}\cos(4\varphi)P_{4,4}\right\}\end{bmatrix}, \quad (39)$$

where $L_{0,0}^{(Ca^2)}$ is included to satisfy the volume conservation condition given in Eq. (32). The expression of $L_{0,0}^{(Ca^2)}$ thus evaluated from the volume conservation constraint is given by

$$L_{0,0}^{(Ca^2)} = -\frac{12}{5}\hat{L}_{2,2}^{(Ca)}. \quad (40)$$

All the other constant coefficients in Eq. (39) are given in Appendix B. Even though the present study is strictly valid for $k \sim O(1)$, we obtain the results of Vlahovska et al. [21] for a surfactant laden droplet suspended in a simple shear flow as $k \to \infty$.

The deformation parameter, $D_{fl}$, for the present case can be written as

$$D_{fl} = \frac{\max\{r_s(\theta, \varphi = 0)\} - \min\{r_s(\theta, \varphi = 0)\}}{\max\{r_s(\theta, \varphi = 0)\} + \min\{r_s(\theta, \varphi = 0)\}}, \quad (41)$$

where $D_{fl}$ gives a measure of the deformation of the droplet in the plane of shear, when suspended in a simple shear flow.

The steady state angle of inclination of the droplet for $\theta = \pi/2$ is given by

$$\varphi_d = \frac{\pi}{4} - \frac{L_{2,2}^{(1)}}{2\hat{L}_{2,2}^{(0)}} Ca + O(Ca^2) \quad (42)$$

The above expression is obtained by performing a Taylor series expansion about $\varphi = \pi/4$. Another alternative method to calculate the inclination angle, $\varphi_d$, for a given value of $\theta$ is to find the value of $\varphi$ corresponding to the maximum value of $r_s$.

## IV. SUSPENSION RHEOLOGY

Next we move on to the calculation of the suspension rheology of a dilute emulsion of droplets suspended in an imposed flow which may be a uniaxial extensional flow or a simple shear flow. According to Batchelor [45], the volume averaged suspension stress for a suspension of force-free particles in linear flow is given by

$$\langle \boldsymbol{\tau} \rangle = -\langle p \rangle \mathbf{I} + 2\mathbf{D}_\infty + \frac{\phi}{V_d}\mathbf{S}, \quad (43)$$

where $\mathbf{S}$ is a stresslet, which signifies the change in total stress as a result of change in velocity and stress due to the presence of a droplet in the flow field. It can be expressed in the following manner [45]

$$\mathbf{S} = \int_{\varphi=0}^{2\pi}\int_{\theta=0}^{\pi} \left[\frac{1}{2}\{(\boldsymbol{\tau}\cdot\mathbf{n})\mathbf{x} + ((\boldsymbol{\tau}\cdot\mathbf{n})\mathbf{x})^T\} - \frac{1}{3}\mathbf{I}\{(\boldsymbol{\tau}\cdot\mathbf{n})\cdot\mathbf{x}\} - \{\mathbf{un} + (\mathbf{un})^T\}\right]d\theta d\varphi. \quad (44)$$

For the case of extensional flow in the far-field, the Trouton or the effective extensional viscosity of a dilute emulsion of droplets is given by

$$\frac{\mu_{ext}}{\mu_e} = \langle \tau_{33} \rangle - \langle \tau_{22} \rangle = \langle \tau_{33} \rangle - \langle \tau_{11} \rangle$$

$$= 3\left[1 + \frac{5}{2}\phi\left\{\begin{array}{l}\dfrac{(-5\lambda + k - 2)\beta + 5\lambda + 2}{(k - 5 - 5\lambda)\beta + 5 + 5\lambda} \\ \left(m_0^{(1)}\lambda^3 + m_1^{(1)}\lambda^2 + m_2^{(1)}\lambda + m_3^{(1)}\right)\beta^3 + \left(m_0^{(2)}\lambda^3 + m_1^{(2)}\lambda^2 + m_2^{(2)}\lambda + m_3^{(2)}\right)\beta^2 \\ +15\dfrac{+\left(m_0^{(3)}\lambda^3 + m_1^{(3)}\lambda^2 + m_2^{(3)}\lambda + m_3^{(3)}\right)\beta + (19\lambda + 16)(25\lambda^2 + 41\lambda + 4)}{56(\beta k - 5\beta - 5\beta\lambda + 5 + 5\lambda)^3}Ca\end{array}\right\}\right], \quad (45)$$

where the different constants present in the above expression are given in Appendix C.

When a simple shear flow is imposed in the suspending fluid, the effective shear viscosity of the dilute emulsion is given by

$$\frac{\mu_{eff}}{\mu_e} = \langle \tau_{12} \rangle = 1 + \frac{5}{2}\left\{\frac{(-5\lambda + k - 2)\beta + 5\lambda + 2}{(-5 - 5\lambda + k)\beta + 5 + 5\lambda}\right\}\phi + O(Ca^2). \quad (46)$$

As can be seen from the above expression, there is no $O(Ca)$ contribution to the effective shear viscosity. The first and second normal stress differences $(N_1$ and $N_2)$ are given by

$$N_1 = \langle \hat{\tau}_{11} \rangle - \langle \hat{\tau}_{22} \rangle$$
$$= 5\frac{\left(n_{1,0}^{(1)}\lambda^2 + n_{1,1}^{(1)}\lambda + n_{1,2}^{(1)}2\right)\beta^2 + \left(n_{2,0}^{(1)}\lambda^2 + n_{2,1}^{(1)}\lambda + n_{2,2}^{(1)}\right)\beta + (16 + 19\lambda)^2}{8(-5\beta - 5\lambda\beta + k\beta + 5 + 5\lambda)^2}\phi Ca,$$

$$N_2 = \langle \hat{\tau}_{22} \rangle - \langle \hat{\tau}_{33} \rangle$$
$$= -5\left[\begin{array}{l}\left(n_{1,0}^{(2)}\lambda^3 + n_{1,1}^{(2)}\lambda^2 + n_{1,2}^{(2)}\lambda + n_{1,3}^{(2)}\right)\beta^3 + \left(n_{2,0}^{(2)}\lambda^3 + n_{2,1}^{(2)}\lambda^2 + n_{2,2}^{(2)}\lambda + n_{2,3}^{(2)}\right)\beta^2 \\ +\dfrac{\left(n_{3,0}^{(2)}\lambda^3 + n_{3,1}^{(2)}\lambda^2 + n_{3,1}^{(2)}\lambda + n_{3,3}^{(2)}\right)\beta + 10(19\lambda + 16)(29\lambda^2 + 61\lambda + 50)}{112(k\beta - 5\lambda\beta - 5\beta + 5 + 5\lambda)^3}\end{array}\right]\phi Ca \quad (47)$$

where the different constants are provided in Appendix C. We again obtain the results of Vlahovska et al. for bulk rheology in the limiting case of $k \to \infty$ although our theory considers a finite value of $k(\sim 1)$.

## V. RESULTS AND DISCUSSIONS

### A. Droplet deformation

#### 1. Uniaxial extensional flow

We first provide a validation for our theoretical results with the numerical results obtained by Milliken et al. [22]. The variation of deformation parameter with $Ca$ is first shown in Fig. 3 for three different values of $\lambda$ $(\lambda = 0.1, 1, 10)$ and for each case the results are validated with that obtained by Milliken et al. [22] The other parameters used in the plot are $\beta = 0.5$ and $k = 0.1$. As can be seen from the Fig. 3, there is a good match between our $O(Ca^2)$ theory and the numerical results as obtained by Milliken et al. [22]. In both the cases, the droplet deformation increases with $Ca$. The $O(Ca)$ theory (first developed by Stone and Leal [15]), however, deviates from the numerical results at a much earlier point. It is also seen that the match is much better for the case of a low value of $\lambda (=0.1)$ as compared to a highly viscous droplet $(\lambda = 10)$. It is observed from Fig. 3 that for a fixed value of $Ca$, the droplet deformation reduces with decrease in $\lambda$. This result is similar to the case of a surfactant-free droplet. The effect of viscosity ratio, $\lambda$, on the deformation of a surfactant-free droplet has been investigated previously by Bentley and Leal [46].

The effect of surfactants on the deformation of the droplet provides us some interesting results. A physical insight on the effect of surfactants as well as viscosity ratio on the deformation of the droplet can be obtained with the help of a contour plot as shown in Fig. 4. The parameter $k$, which is the property parameter, when increased, enhances the convection of surfactants along the surface of the droplet. Due to the imposed extensional flow the surfactants start accumulating at the two tips of the droplet along the z-axis. In addition to this, if $k$ is increased, the concentration of the surfactants further increases at the tip due to increase in convection. This results in a lower surface tension at the tips as compared to the other regions on its surface. In other words, the surface tension gradient along the surface increases with increase in $k$ and hence the surfactant-induced Marangoni stress increases, which causes a larger deformation of the droplet as compared to a clean droplet. Also accumulation of surfactants at the end of the droplet along the axis of elongation of the droplet, requires a higher curvature and hence results in an increased deformation.

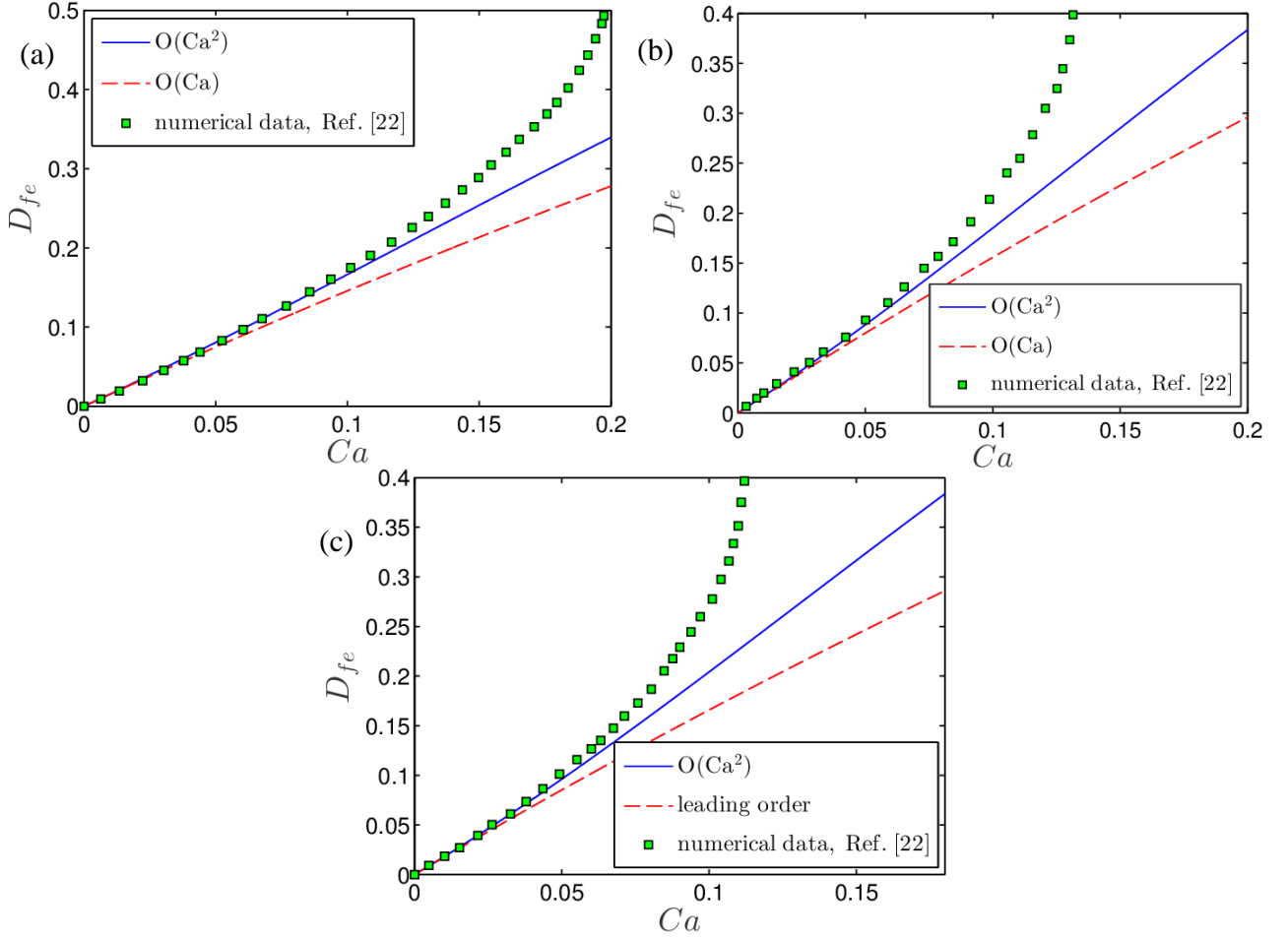

Fig. 3. Variation of deformation parameter $(D_{fe})$ with $Ca$ is shown. For Fig. (a) $\lambda = 0.1$, Fig. (b) $\lambda = 1$ and (c) $\lambda = 10$. In each of these figures numerical data from the work done by Milliken et al. is shown along with $O(Ca)$ and $O(Ca^2)$ solutions obtained from our theory. The value of other parameter in these plots are $\beta = 0.5,$ and $k = 0.1$.

It is confirmed from Fig. 4 that for a particular value of $\beta$, increase in $k$ increases the deformation of the droplet. For a droplet having fixed concentration of surfactants along its surface (that is $\beta = 0$), the average surfactant concentration decreases upon deformation. Thus the surface tension increases and any further deformation of the droplet reduces in comparison to a droplet having uniform surfactant distribution.

If $\lambda$ is increased, or if the droplet is highly viscous as compared to the suspending flow, then any change in the surface tension on the surface of the droplet (and hence the Marangoni stress) does not affect its deformation to that extent as it would had been for a low viscous droplet. In other words, the reduction in surface velocity of the droplet, due to increase in Marangoni stress, doesn't contribute much to the deformation characteristics for a high viscous droplet. This is because the surface velocity is already significantly reduced due to the high viscosity of the droplet and any further reduction in the same due to variation in surfactant concentration is just incremental. This can as well be observed by comparing Fig 4(a) and 4(b).

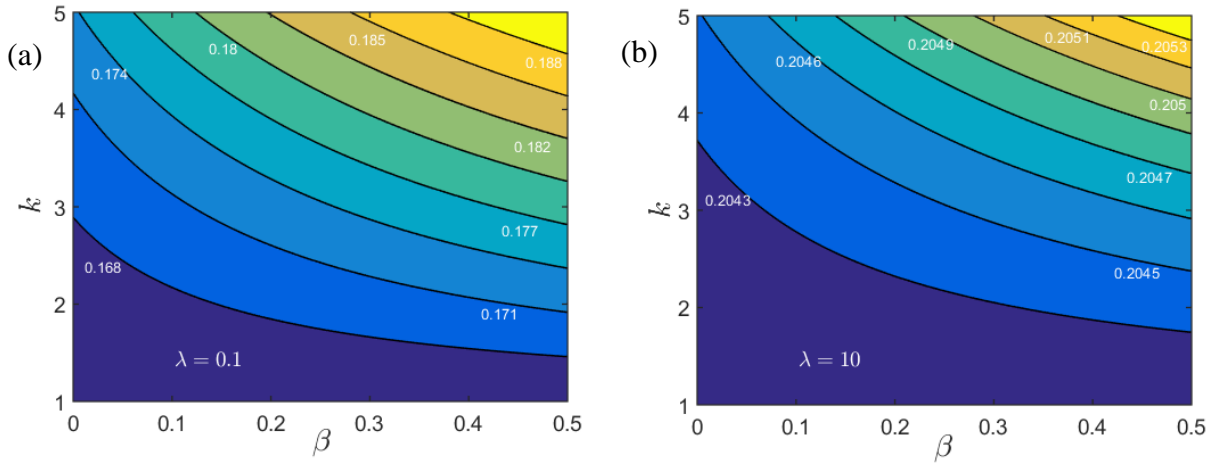

Fig. 4. Contour plot showing the variation of deformation parameter $(D_{fe})$ with $\beta$ and $k$. For Fig. (a) $\lambda = 0.1$ while in Fig. (b) $\lambda = 10$. The values of the deformation parameter corresponding to different values of $\beta$ and $k$ are also provided in the plot above. The value of capillary number for the above plot is taken to be $Ca = 0.1$.

The change in deformation parameter $(D_{fe})$ of a droplet with $\lambda = 10$ due to change in $\beta$ or $k$ is found to be much less in comparison to that of a droplet with $\lambda = 0.1$. It can also be said that for a high value of $k$ $(e.g.\ k = 5)$, the Marangoni stress is high enough to affect the deformation of the droplet and any change in $\lambda$ (say from $\lambda = 0.1$ to $\lambda = 1$) has minimal effect on the same. For smaller values of $k$, the Marangoni stress developed is low and hence any change in $\lambda$ has significant effect on the deformation of the droplet. At the same time, from Fig. 4 we can also say that an increase in the viscosity ratio (from $\lambda = 0.1$ to $\lambda = 10$) reduces the effect of surfactants on the deformation of the droplet. For $\lambda = 0.1$, the deformation parameter varies from a minimum of 0.168 to a maximum of 0.188, whereas for $\lambda = 10$, the change in $D_{fe}$ is way too small, that is from 0.2043 to 0.2053.

The deformed shape of the droplet, subjected to a uniaxial extensional flow is shown in Fig. 5 for different orders of perturbation, $O(Ca)$ and $O(Ca^2)$. It is seen that due to the presence of extensional bulk flow the droplet takes the shape of an ellipsoid with its major axis aligned along the extensional axis. Higher order correction shows that the droplet becomes more ellipsoidal.

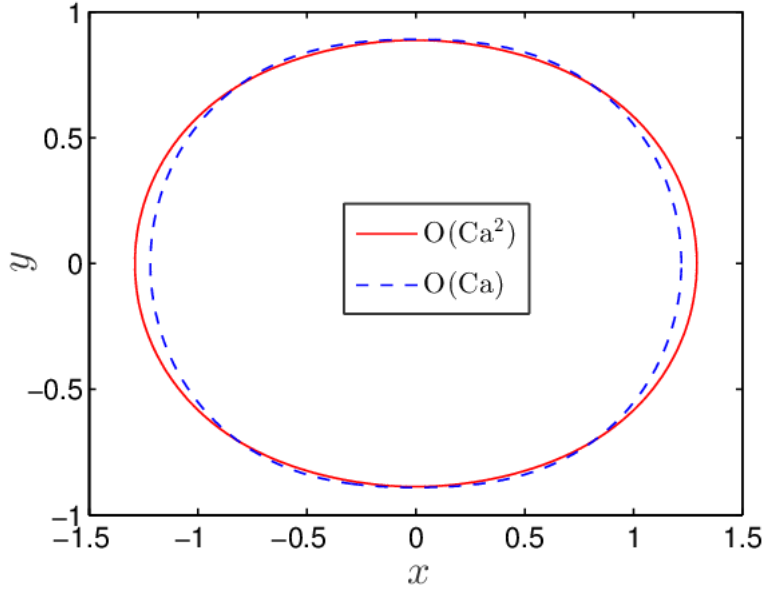

Fig. 5. Deformed shape of the droplet at different orders of perturbation. The different parameters involved in this plot are $\beta = 0.5$, $k = 0.1$, $\lambda = 1$, and $Ca = 0.1$.

## 2. Simple shear flow

We first validate our result with existing experimental results of Feigl et al. [27] Towards this, we first plot the variation of the parameters $L, B$ and $W$ with $Ca$. These parameters have been previously used by Feigl et al. [27] to define different experimentally observed droplet dimensions. $L$ denotes the dimensionless major axis of the deformed droplet, which increases as the surface tension force reduces in comparison to the viscous forces acting on the droplet (or as $Ca$ increases). $B$ indicates the minor axis of the droplet and it reduces with increase in $Ca$. Finally $W$ is the length along the vorticity axis which too reduces with increase in $Ca$. These parameters thus give a measure of the deformation of the droplet and can be expressed as

$$\left. \begin{array}{l} L = \max_{\varphi}\left\{ r_s\left(\theta = \pi/2, \varphi \in [0,\pi]\right)\right\}, \\ B = \min_{\varphi}\left\{ r_s\left(\theta = \pi/2, \varphi \in [0,\pi]\right)\right\}, \\ W = \min_{\theta}\left\{ r_s\left(\varphi = \pi/2, \theta \in [0,\pi/2]\right)\right\}. \end{array} \right\} \qquad (48)$$

In Fig. 6 we see that for all the three parameters $(L, B, W)$, there is a good match between the experimental results of Fiegl et al. [27] and our $O(Ca^2)$ solution. The $O(Ca)$ solution, however, largely deviates from the experimental result in comparison to a higher order solution. Due to the presence of bulk shear flow, the surfactants accumulate on either of the tips of the major axis, while there is a dearth of the same at the end of both the minor as well as the vorticity axis. Due to this non-uniform distribution of surfactants along the droplet surface, a gradient in surface tension is generated which results in Marangoni stress. This results in the deformation of the droplet.

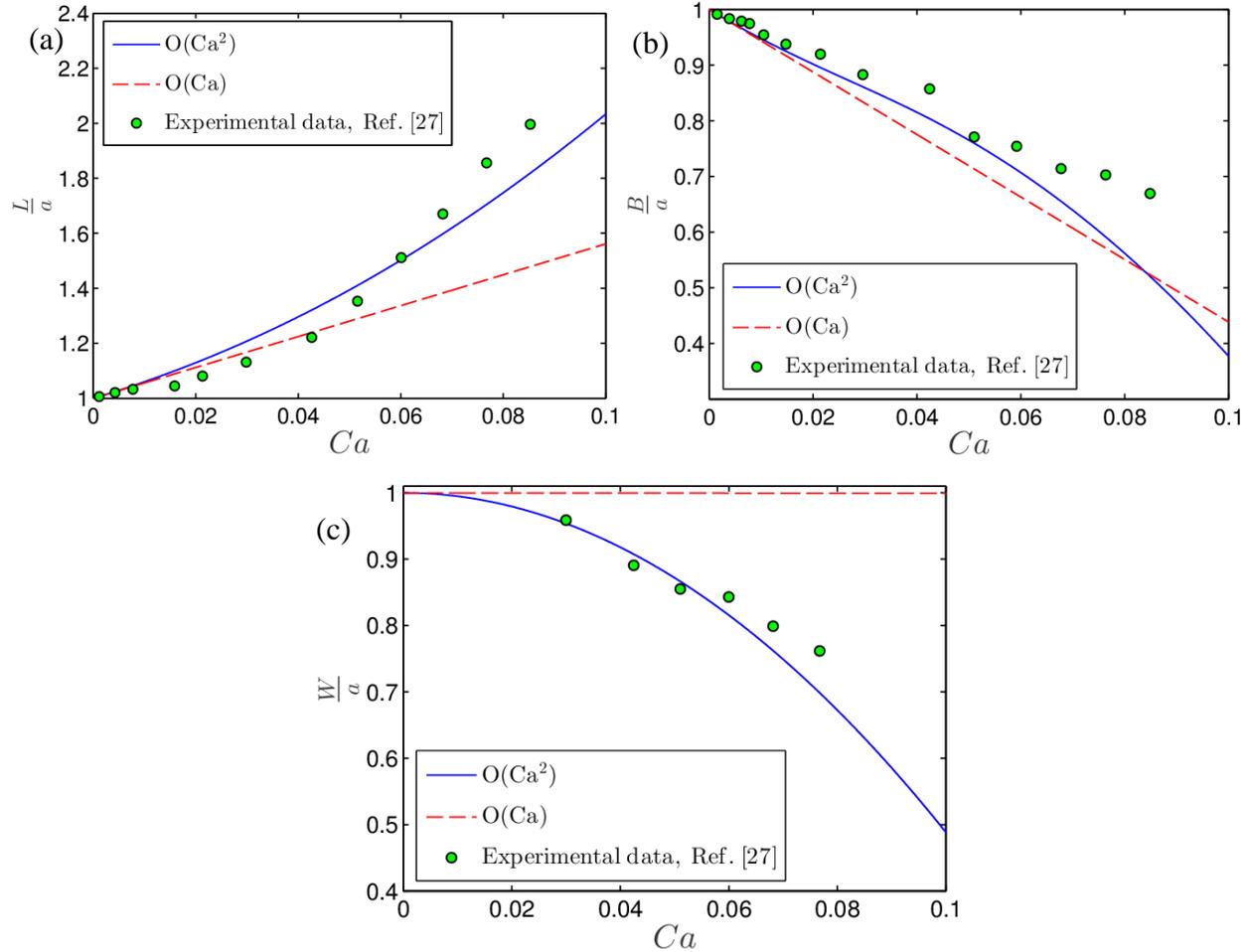

Fig. 6. Variation of $O(Ca)$ and $O(Ca^2)$ solution for (a) *L*, (b) *B*, and (c) *W* with *Ca*. The circular points indicate the experimental data point as obtained from Fig. 4 of Fiegl et.al. [27] The values of the other parameters are $\beta = 0.8, \lambda = 0.335$ and $k = 1$.

We next explore the effect of surfactant distribution as well as the viscosity ratio on the deformation of the droplet. Towards this, the variation of the deformation parameter, $D_{fl}$, with *Ca*, as obtained from both $O(Ca)$ and $O(Ca^2)$ solutions, is shown in Fig. 7. The variation of the deformation parameter with *Ca* as obtained by Feigl et al. [27] is also shown in the same plot. As can be seen from Fig. 7. that there is a good match between our theoretical prediction and the experimental results of Feigl. et al. The plot shown in Fig. 7 indicates that the droplet deformation increases with increase in *Ca* .

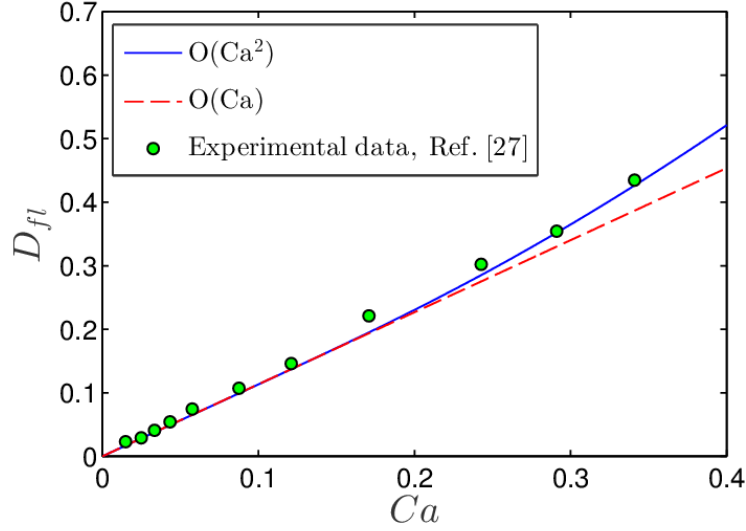

Fig. 7. Variation of $O(Ca)$ and $O(Ca^2)$ solution for $D_{fl}$ with $Ca$. The circular points indicate the experimental data points as obtained from the paper of Feigl et.al. The values of the other parameters are $\beta = 0.5, \lambda = 0.335$ and $k = 5$.

Towards investigating the effect of surfactant distribution on the deformation of the droplet, we show two contour plots in Fig. 8. Fig. 8(a) considers the case of a low viscous droplet with $\lambda = 0.1$ and Fig. 8(b) considers the case of a highly viscous droplet with $\lambda = 10$. It can be seen from the contour plot in Fig. 8 that increase in both $\beta$ and $k$ increases the deformation of the droplet. Increase in $k$ increases the non-uniformity in surfactant distribution due to enhanced surfactant convection. Hence the surface tension gradient along the droplet surface increases which further results in a higher Marangoni stress. This Marangoni stress developed due to non-uniform surfactant distribution is responsible for the droplet deformation. For low viscous droplets $(\lambda = 0.1)$, the Marangoni stress plays an important role in the deformation of the droplet. For high viscous droplets $(\lambda = 10)$, although the deformation of the droplet for same values of $\beta$ and $k$ is larger, the effect of Marangoni stress on droplet deformation is minimal. This can be observed by comparison of Fig. 8(a) and Fig. 8(b). It is thus seen that increase in droplet deformation due to increase in either $\beta$ or $k$ is larger for $\lambda = 0.1$ as compared to the case when $\lambda = 10$. Thus it can be said that change in surfactant distribution along the surface of the droplet results in a significant change in deformation for a bubble $(\lambda \to 0)$, where no deformation takes place for a particle $(\lambda \to \infty)$.

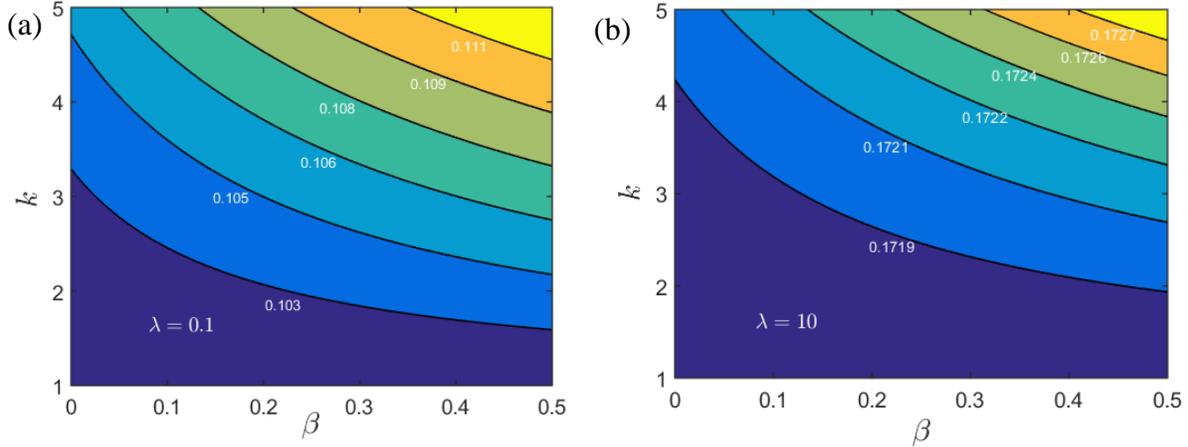

Fig. 8. Contour plot showing the variation of deformation parameter $(D_{fl})$ with $\beta$ and $k$. For Fig. (a) $\lambda = 0.1$ while in Fig. (b) $\lambda = 10$. The values of the deformation parameter corresponding to different values of $\beta$ and $k$ are labeled in the plot above. The value of capillary number for the above plot is taken to be $Ca = 0.1$.

The angle of inclination can be also theoretically predicted. For $\theta = \pi/2$, the expression for angle of inclination as given in Eq. (42), when plotted against Ca, matches pretty well with the numerical result obtained by Li and Pozrikidis [23]. This is shown in Fig. 9(a).

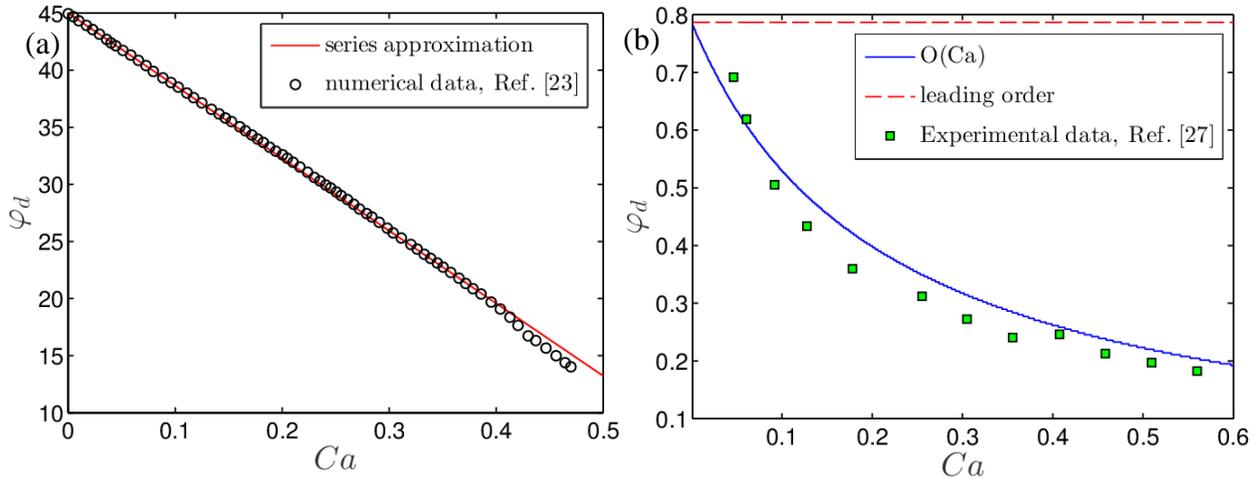

Fig. 9. Variation of inclination angle with $Ca$. (a) a series approximation for the inclination angle is found out (Eq. (42)) and plotted against $Ca$. The different parameters involved are $k = 10, \lambda = 1$ and $\beta = 0.1$. (b) a more accurate approach is used to calculate the angle of inclination directly from the solution for the deformed droplet shape. The 'green' square points denote the experimental data points from the work done by Feigl et.al. The other parameters are $\lambda = 6.338, \beta = 0.5$ and $k = 5$.

The angle of inclination for any particular value of $\theta$ can, however, be found out directly from the maximum dimension of the droplet, $(L)$. The angle of inclination, thus found out, in such a manner is seen to decrease with increase in *Ca*, that is, the droplet is found to orient itself in the direction of the flow. Also a good match is observed between the experimental results of Feigl et al. [27] and $O(Ca)$ solution to the inclination angle (see Fig. 9(b)). The leading order solution for $\varphi_d$ has no variation with *Ca*.

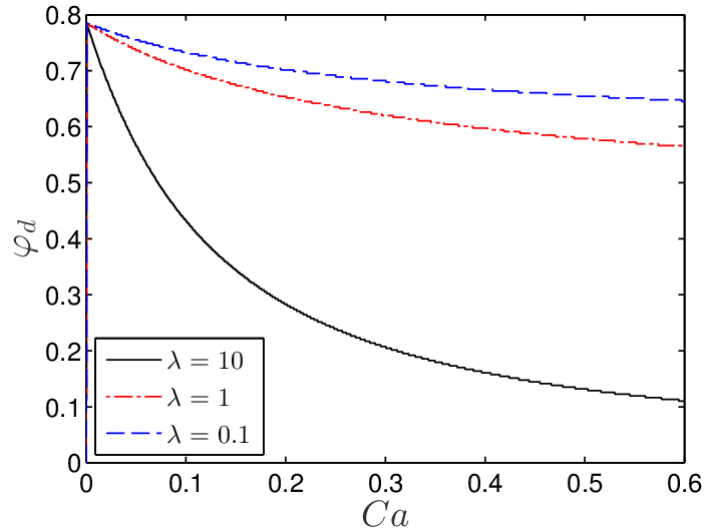

Fig. 10. Variation of the inclination angle of the droplet with *Ca* for different values of $\lambda (= 0.1, 1, 10)$. The other parameters values are $\beta = 0.5$, $k = 5$, $\lambda = 1$ and $\theta = \pi/2$.

The effect of droplet viscosity or the viscosity ratio, $\lambda$, on the inclination angle of the droplet can be explained from Fig.10. It is seen that higher the viscosity ratio, lower is the inclination angle for a given value of capillary number. In other words, a droplet with a higher viscosity aligns itself more towards the direction of imposed flow in comparison to a low viscous droplet. It is also seen that surfactant distribution along the droplet surface has no effect on the orientation or the inclination angle of the droplet. Fig. 10 is drawn for a fixed value of the polar angle, $\theta = \pi/2$. For higher *Ca*, change in $\lambda$ has a greater effect on the orientation of the droplet suspended in a simple shear flow.

The shape of a surfactant laden droplet suspended in a simple shear flow is shown for different orders of perturbation in Fig. 11. This figure clearly shows the orientation of the droplet as well. The droplet is found to elongate and take the shape of an ellipsoid with the major axis oriented along the extensional axis. The surfactants thus accumulate at the either ends of the major axis and affect the droplet deformation.

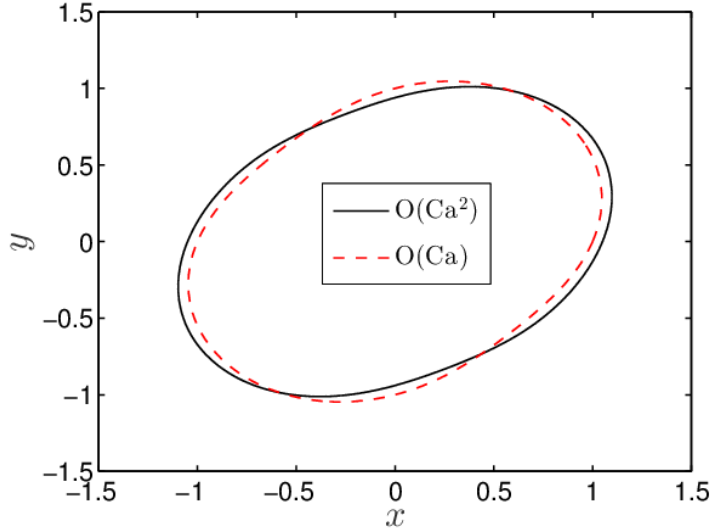

Fig. 11. Shape of the surfactant laden-droplet when suspended in a simple shear flow. Both the shapes due to $O(Ca)$ as well as $O(Ca^2)$ corrections are shown. The different parameters used for this plot are $\beta = 0.1$, $k = 5$, $\lambda = 1$ and $Ca = 0.15$.

### B. Suspension rheology

#### 1. Uniaxial extensional flow

In Fig. 12 we show the variation of normalized Trouton viscosity or the effective extensional viscosity with the bulk viscosity ratio, $\lambda$. The effective extensional viscosity is normalized with respect to the volume fraction, $\phi$. For the limiting case of a clean droplet $k = 0$, our theoretical result matches exactly with that obtained by Ramachandran and Leal [45] (they have considered the effect of slip in their analysis, however, we have taken the slip factor to be zero in order to match our result with theirs). For the case of clean droplets $(k = 0)$, the presence of the droplets in the suspending fluid tends to retard the imposed flow. This results in a viscosity of the emulsion which is larger than the viscosity of the bulk fluid. As the viscosity of the surfactant-free droplet increases (or as $\lambda$ increases), the resistance provided by the suspended droplet increases and hence the effective extensional viscosity $(\mu_{ext})$ increases too (see Fig. 12). In the limiting case of particle $(\lambda \to \infty)$, the effective extensional viscosity is the highest.

Presence of surfactants on the surface of the droplets further modifies the effective extensional viscosity. Non-uniform distribution of the surfactants induced by the bulk flow, generates a Marangoni stress due to variation of surface tension about the droplet surface. This Marangoni stress, which acts against the direction of bulk flow deforms the droplet and further increases the effective extensional viscosity of the droplet. Increase in $k$ increases the convective transport of surfactants along the droplet surface and hence increases the non-uniformity in surfactant distribution, which in turn increases the Marangoni stress. As a result, $\mu_{ext}$ increases with increase in $k$. This can clearly be seen from Fig. 12(a). Another important observation from the same figure is that the increase in $\mu_{ext}$ with increase in $k$ is the largest for the case of low viscous droplets. That is, the surfactant concentration on droplet surface has almost negligible effect on the effective extensional viscosity for a highly viscous droplet, as the effect of Marangoni stress is minimal. Also comparison of Fig. 12(a) and Fig. 12(b) shows us the effect of the parameter $\beta$ on the bulk rheology. The elasticity parameter, $\beta$, increases the sensitivity of surface tension towards the surfactant distribution and hence for a constant $k$, increase in $\beta$ increases the Marangoni stress, which in turn increases the effective extensional viscosity. That is, the effect of $k$ on bulk rheology is enhanced with increase in $\beta$.

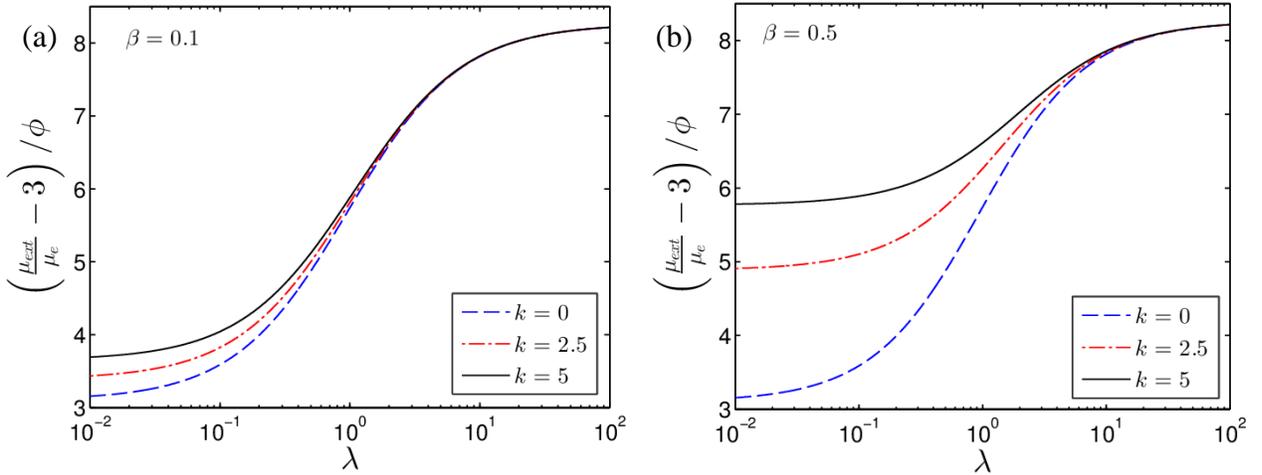

Fig. 12. Variation of normalized effective extensional viscosity with $\lambda$ for different values of $k\ (=0, 2.5, 5)$. Variation of the normalized effective extensional viscosity with $\lambda$ for a clean droplet $(k=0)$ has been shown by Ramachandran and Leal. Fig. (a) shows the results for $\beta = 0.1$ while Fig. (b) is for $\beta = 0.5$. The value of the capillary number is taken as $Ca = 0.1$.

### 2. Simple shear flow

Next a variation of the effective shear viscosity $\left(\mu_{eff}\right)$ of a dilute emulsion of droplets suspended in a simple shear flow with $\lambda$ for different values of $k(=0, 2.5, 5)$ is shown in Fig. 13(a). The nature of variation of the effective shear viscosity is the same as was observed for the case of effective extensional viscosity. The effective shear viscosity $\left(\mu_{eff}\right)$ is seen to increase with increase in non-uniformity of surfactants along the droplet surface That is, higher Marangoni stress acting on the droplets (large $k$) results in higher $\mu_{eff}$. This effect of surfactant distribution on $\mu_{eff}$ is seen mainly for low viscous droplets. Comparison of Fig. 13(a) and Fig.

13(b) shows that increase in $\beta$ increases the effect of $k$ on the effective shear viscosity, in a similar manner as seen for the case of a uniaxial extensional bulk flow.

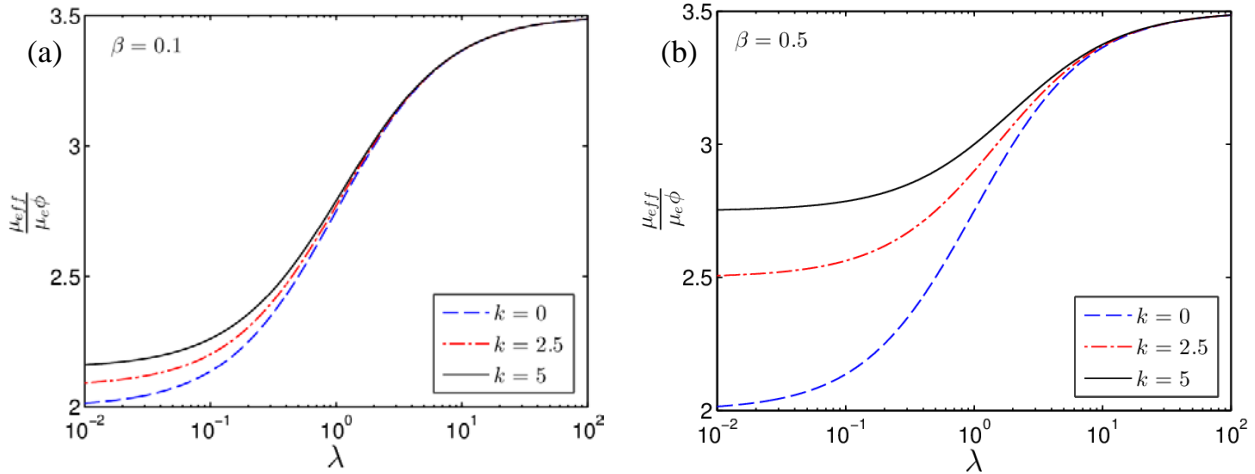

Fig. 13. Variation of normalized effective shear viscosity with $\lambda$ for different values of $k$ $(k=0, 2.5, 5)$. Fig. (a) considers the case for $\beta = 0.1$, while Fig. (b) is plotted for $\beta = 0.5$. The value of the capillary number is $Ca = 0.1$.

The deformation of the droplets suspended in a simple shear flow essentially depends on the imposed shear rate. Thus a dilute emulsion of droplets necessarily shows non-Newtonian behavior [47]. As a result, they also exhibit properties like normal stresses $(\tau_{11}, \tau_{22}, \tau_{33})$ which can be seen in fluids that display elastic behavior. The capillary number, $Ca$, which is dependent on the imposed shear rate, is a governing factor for the deformation of the droplet. It has been shown previously, that deformation of droplets in the emulsion actually results in the generation of normal stresses [47]. This is because the model of an emulsion of two Newtonian fluids (droplet and the carrier phase) is seen to have elastic properties as present in various complex fluids. As suspension of droplets in a simple shear flow results in non-isotropic normal stresses, difference between the different components of normal stresses exists. These normal stresses exerted by the suspension in shear flow is different in each direction, so that $N_1$ and $N_2$ are non-zero.

The variation of the first and second normal stress differences, $N_1$ and $N_2$, with $Ca$ is shown in Fig. 14. For both the cases, there is a good match between our theoretical prediction and the numerical results of Li and Pozrikidis [23]. Both the magnitude of $N_1$ and $N_2$ increase with increase in $Ca$. Under the assumption of negligible inertia, $N_1$ is positive and $N_2$ is negative for a dilute emulsion of droplets. This can be seen from Fig. 14 as well as in Fig. 15. The sign of these normal stress differences is related to the deformation of a droplet suspended in a simple shear flow. As seen from Fig 11, at $O(Ca)$ the initially spherical droplet is stretched into an ellipsoidal shape with the major axis aligned along the extensional axis of the simple shear.

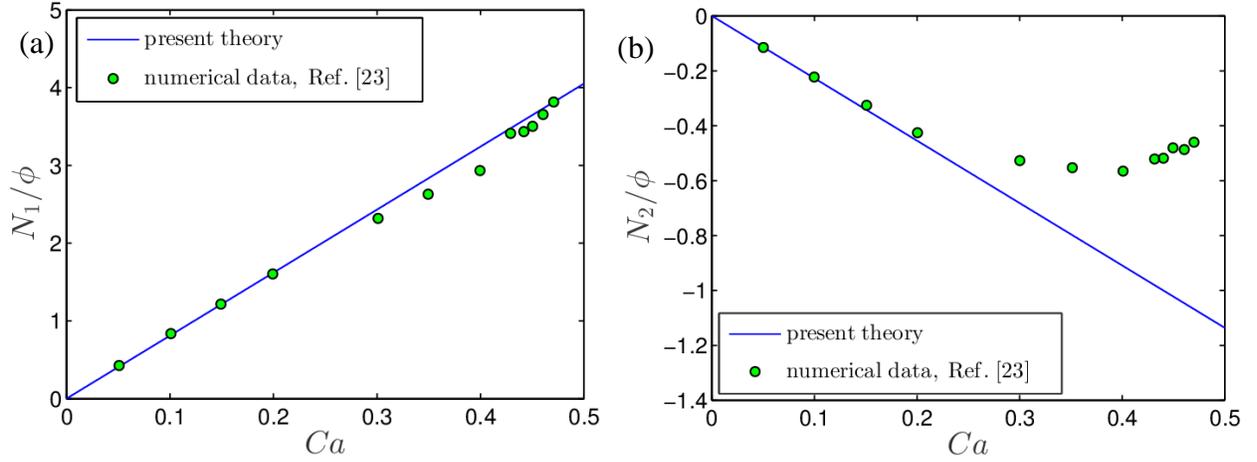

Fig. 14. (a) Variation of first normal stress difference $(N_1)$ with $Ca$. (b) Variation of the second normal stress difference $(N_2)$ with $Ca$. The circular points in the plot indicate the numerical result as obtained by Li and Pozrikidis [23]. The different parameters involved in either of the plots are $\beta = 0.1$, $k = 10$ and $\lambda = 1$.

Finally we show the variation of $N_1$ and $N_2$ with $\lambda$ for different values of $k\,(=0, 2.5, 5)$ in Fig. 15. It can be seen from Fig. 15 that both the normal stresses increase with an increase in $\lambda$. There is also an increase in $N_1$ due to an increase in $k$. This increase is the largest for the case of a low viscous droplet as compared to a highly viscous droplet. At $O(Ca^2)$, the shape of the droplet is unaffected by rotation and is directly proportional to the rate of strain tensor. However, the vorticity present in the simple shear bulk flow tends to rotate the ellipsoidal droplet in the flow direction. The tensile component of surface tension forces that act in this direction, thus results in a positive $N_1$ at $O(Ca)$, whereas the extra compressive stress acting on the droplet in the gradient direction causes a negative $N_2$ for the same order.

To investigate the effect of surfactant concentration on $N_1$ and $N_2$, we have shown the variation of the normalized first and second normal stress difference $(N_1/\phi, N_2/\phi)$ with the viscosity ratio, $\lambda$ for different values of $k$. It can be seen from Fig. 15(a) that $N_1$ increases as the droplet phase is made more viscous or in other words, as $\lambda$ is increased. For the case of a surfactant-laden droplet the surface tension along the droplet interface varies due to the non-uniform distribution of surfactants. Thus with increase in $k$, the gradient in surface tension along the droplet surface increases. This results in an increase in the tensile component of the surface tension force along the flow direction thus elongating the droplet along the extensional axis. This in turn results in an increase in the magnitude $N_1$ which can be seen from Fig. 15(a). The effect of surfactant distribution on $N_1$ is seen to be higher for a low viscous droplet as compared to a highly viscous droplet.

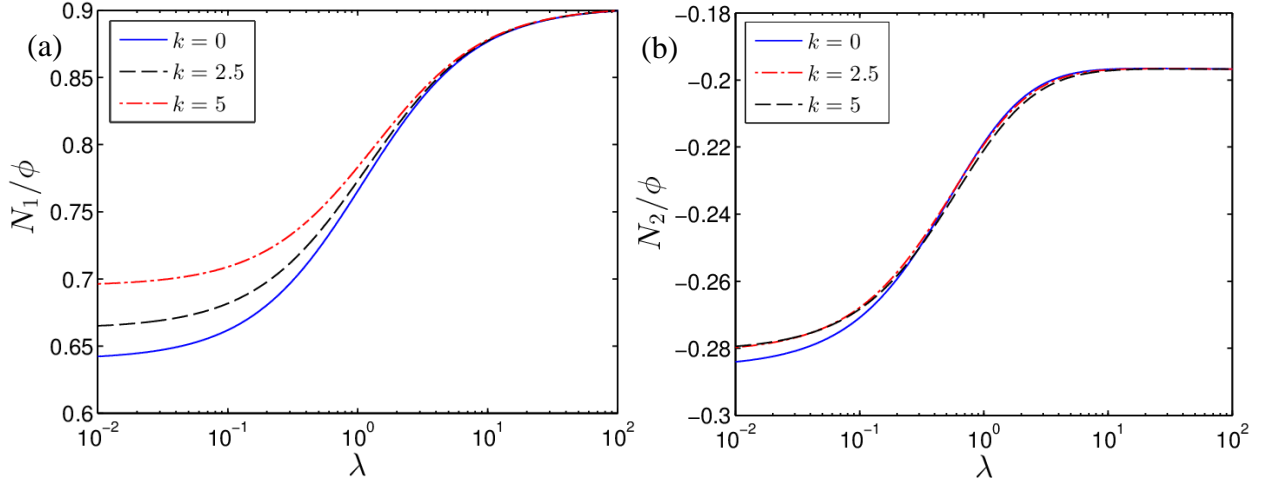

Fig. 15. (a) Variation of normalized first normal stress difference $(N_1/\phi)$ with $\lambda$. (b) Variation of the normalized second normal stress difference $(N_2/\phi)$ with $\lambda$. Each of the plots are drawn for different values of $k(=0, 2.5, 5)$. The different parameters involved in either of the plots are $\beta = 0.1$ and $\lambda = 1$.

Now looking into Fig. 15(b), we can find that the nature of variation of $N_2$ is just the opposite as compared to $N_1$. As $\lambda$ increases, the magnitude of $N_2$ decreases. For a low viscous droplet, if compare between a clean droplet and surfactant laden droplet, we can say that for the later the Marangoni stress generated due to non-uniform distribution of surfactants results in a compression along the gradient direction. This results in a decrease of $N_2$ for a surfactant-laden droplet. However, change in surfactant concentration or any variation of the parameter $k$ doesn't have any effect on $N_2$ as seen from Fig. 15(b). Just like in the case of $N_1$, variation in $N_2$ due to presence of surfactants is found to be significant for a low viscous droplet as compared to a highly viscous droplet.

## VI. CONCLUSIONS

In the present study, we have investigated the effect of surfactant distribution on droplet deformation as well as on the bulk rheology of a dilute emulsion of droplets. An asymptotic approach is used to analyze the problem for the limiting case when the surfactant transport along the droplet surface is dominated by the surface diffusion rather than surface convection. A regular perturbation methodology was used, with $Ca$ as the perturbation parameter to solve the flow field in the small deformation limit. The present theory developed matches well with the existing numerical as well as experimental results. The results thus obtained revealed various interesting outcomes, some which are stated below.

(i) For a droplet suspended in a linear flow (uniaxial extensional or simple shear flow), increase in $k$ or $\beta$, enhances the deformation of the droplet. This effect of surfactant concentration on the droplet deformation reduces as the droplet becomes more viscous or as $\lambda$ increases.

(ii) For the case when the droplet is suspended in a simple shear flow, the inclination angle remains unaffected by any variation in the surfactant concentration along the droplet surface. However, if the viscosity of the droplet with respect to the suspending fluid is increased gradually, the droplet starts aligning itself towards the direction of flow.

(iii) Considering a dilute emulsion of droplets suspended in a linear flow field, the effective shear viscosity (effective extensional viscosity for extensional flow) of the emulsion is significantly affected by both the viscosity ratio as well as the surfactant concentration along the droplet surface. That is, increase in both $k$, $\beta$ or $\lambda$ increases the effective shear viscosity of the emulsion, although the effect of the parameter $k$ is more significant for low value of $\lambda$.

(iv) For the case of dilute emulsion of droplets suspended in a simple shear flow, normal stress differences $(N_1, N_2)$ are present. The first normal stress difference $(N_1)$ is found to increase with increase in $k$, whereas the same also is found to increase for a higher value of $\lambda$ provided $k$ is a constant. The second normal stress difference $(N_2)$, on the other hand, seems to be unaffected by any change in the surfactant concentration, although it reduces with increase in $\lambda$.

**Appendix A: Different constants present in the expressions of surfactant concentration and droplet deformation when the bulk flow is a uniaxial extensional flow**

The constant coefficients present in the expression for surfactant concentration as shown in Eqs. (29) and (30), are given below

$$\left.\begin{array}{l} g1_{2,0} = \{76\lambda^2 + (368-15k)\lambda + 256 - 60k + 4k^2\} \\ g2_{2,0} = \{-228\lambda^2 + (-1104+10k)\lambda + -768 + 100k - 4k^2\} \\ g3_{2,0} = \{228\lambda^2 + (1104+25k)\lambda - 20k + 768\} \\ g4_{2,0} = -76\lambda^2 - (20k+368)\lambda - 256 - 20k \end{array}\right\} \quad (A1)$$

and

$$g1_{4,0} = \{608\lambda^2 + (1120 - 187k)\lambda + 512 - 172k + 20k^2\}$$
$$g2_{4,0} = \{-1824\lambda^2 + (-3360 + 338k)\lambda - 1536 + 308k - 20k^2\}$$
$$g3_{4,0} = \{1824\lambda^2 + (3360 - 115k)\lambda - 100k + 1536\}$$
$$g4_{4,0} = -4(\lambda + 1)(152\lambda + 128 + 9k)$$
(A2)

The expression of the constants present in Eqs. (34), (35) and (36) for $O(Ca^2)$ deformation is given by

$$a_0^{(20)} = -285475, \ a_1^{(20)} = (174100k - 664575),$$
$$a_2^{(20)} = (270800k - 478800 - 35300k^2), \ a_3^{(20)} = 97600k - 102400 + 2400k^3 - 27200k^2,$$
$$b_0^{(20)} = 856425, \ b_1^{(20)} = (-348200k + 1993725),$$
$$b_2^{(20)} = (-541600k + 35200k^2 + 1436400), \ b_3^{(20)} = 307200 + 26800k^2 - 195200k,$$
$$c_0^{(20)} = -856425, \ c_1^{(20)} = (174100k - 1993725),$$
$$c_2^{(20)} = (-1436400 + 100k^2 + 270800k), \ c_3^{(20)} = -307200 + 97600k + 400k^2,$$
(A3)

and

$$a_0^{(40)} = -214035, \ a_1^{(40)} = (-581235 + 114600k),$$
$$a_2^{(40)} = (-19710k^2 + 209100k - 524640), \ a_3^{(40)} = -157440 - 18120k^2 + 95040k + 1080k^3,$$
$$b_0^{(40)} = 642105, \ b_1^{(40)} = (-229200k + 1743705),$$
$$b_2^{(40)} = (19650k^2 - 418200k + 1573920), \ b_3^{(40)} = -190080k + 18000k^2 + 472320,$$
$$c_0^{(40)} = -642105, \ c_1^{(40)} = (-1743705 + 114600k),$$
$$c_2^{(40)} = (209100k - 1573920 + 60k^2), \ c_3^{(40)} = -472320 + 95040k + 120k^2.$$
(A4)

The expression of the velocity field both outside and inside the droplet is given below

$$\mathbf{u}_{i,0} = \frac{15}{4}\left\{\frac{r(1-\beta)(1-r^2)}{5+kbt+5\lambda-5bt-5\lambda bt}\right\}(1-3\cos^2\theta)\mathbf{e}_r + \frac{15}{8}\left\{\frac{r(1-\beta)(3-5r^2)}{5+k\beta+5\lambda-5\beta-5\lambda\beta}\right\}\sin(2\theta)\mathbf{e}_\theta,$$

$$\mathbf{u}_{e,0} = \left[\frac{5}{4}\frac{\left\{\left((-5\lambda+k-2)r^2+3\lambda-\frac{3}{5}k\right)bt+(5\lambda+2)r^2-3\lambda\right\}}{r^4\left\{(k-5-5\lambda)bt+5+5\lambda\right\}}(1-3\cos^2\theta)\right]\mathbf{e}_r \qquad (A5)$$
$$+\frac{3}{2}r\cos^2\theta-\frac{1}{2}r$$
$$+\frac{3}{4}\left[\frac{\{(-5\lambda+k)bt+5\lambda\}}{r^4\{(k-5-5\lambda)bt+5+5\lambda\}}-r\right]\sin(2\theta)\mathbf{e}_\theta,$$

The pressure field for either of the phases are written below

$$p_{i,0} = -\left\{\frac{105\lambda(1-\beta)r^2}{(4k-20-20\lambda)bt+20+20\lambda}\right\}(1-3\cos^2\theta),$$
$$p_{e,0} = \frac{5}{2r^3}\left\{\frac{(-5\lambda+k-2)bt+5\lambda+2}{(k-5-5\lambda)bt+5+5\lambda}\right\}(1-3\cos^2\theta), \qquad (A6)$$

## Appendix B: Constants present in the expressions of surfactant concentration and droplet deformation when the bulk flow is a simple shear flow

The expression of the constant coefficients present in Eq. (38), in the expression for surfactant concentration are written below

$$\hat{\Gamma}^{(0)}_{2,2} = \frac{5k}{12}\left\{\frac{1-\beta}{-5\beta-5\lambda\beta+5\lambda+5+k\beta}\right\},$$

$$\Gamma^{(Ca)}_{2,0} = -25k\frac{\{h1_{2,0}\beta^3+h2_{2,0}\beta^2+h3_{2,0}\beta+h4_{2,0}\}}{336(-5\beta-5\beta\lambda+\beta k+5+5\lambda)^3},$$

$$\Gamma^{(Ca)}_{4,0} = -45k\frac{\{h1_{4,0}\beta^3+h2_{4,0}\beta^2+h3_{4,0}\beta+h4_{4,0}\}}{224(\beta k-9\beta-9\beta\lambda+9\lambda+9)(-5\beta-5\beta\lambda+\beta k+5+5\lambda)^2}, \qquad (B1)$$

$$\Gamma^{(Ca)}_{2,2} = 5k\frac{\{h1_{2,2}\beta^2+h2_{2,2}\beta+h3_{2,2}\}}{288(-5\beta-5\beta\lambda+\beta k+5+5\lambda)^2},$$

$$\Gamma^{(Ca)}_{4,4} = 5k\frac{\{h1_{4,4}\beta^3+h2_{4,4}\beta^2+h3_{4,4}\beta+h4_{4,4}\}}{5376(\beta k-9\beta-9\beta\lambda+9\lambda+9)(-5\beta-5\beta\lambda+\beta k+5+5\lambda)^2},$$

where

$$h1_{2,0} = \{76\lambda^2 + (368-15k)\lambda + 256 - 60k + 4k^2\}$$
$$h2_{2,0} = \{-228\lambda^2 + (-1104+10k)\lambda - 768 + 100k - 4k^2\}$$
$$h3_{2,0} = \{228\lambda^2 + (1104+25k)\lambda - 20k + 768\}$$
$$h4_{2,0} = \{-76\lambda^2 + (-20k-368)\lambda - 20k - 256\}$$

(B2)

$$h1_{4,0} = \{608\lambda^2 + (1120-187k)\lambda + 512 - 172k + 20k^2\}$$
$$h2_{4,0} = \{-1824\lambda^2 + (-3360+338k)\lambda - 1536 + 308k - 20k^2\}$$
$$h3_{4,0} = \{1824\lambda^2 + (3360-115k)\lambda - 100k + 1536\}$$
$$h4_{4,0} = -4(\lambda+1)(152\lambda + 9k + 128)$$

(B3)

$$h1_{2,2} = 19\lambda^2 + (92+16k)\lambda + 64 + 4k$$
$$h2_{2,2} = -38\lambda^2 - (184+36k)\lambda - 128 - 24k$$
$$h3_{2,2} = 19\lambda^2 + (92+20k)\lambda + 64 + 20k$$

(B4)

and

$$h1_{4,4} = \{608\lambda^2 + (1120-187k)\lambda + 512 - 172k + 20k^2\},$$
$$h2_{4,4} = \{-1824\lambda^2 + (-3360+338k)\lambda - 1536 + 308k - 20k^2\},$$
$$h3_{4,4} = \{1824\lambda^2 + (3360-115k)\lambda - 100k + 1536\},$$
$$h4_{4,4} = -4(\lambda+1)(152\lambda + 9k + 128)$$

(B5)

The constant coefficients in the expression of droplet shape as given in Eq. (39), are given by

$$\hat{L}_{2,2}^{(Ca)} = \frac{5}{48}\left(\frac{16+19\lambda-19\lambda\beta+4k\beta-16\beta}{-5\beta-5\lambda\beta+5\lambda+5+k\beta}\right).$$

(B6)

$$L_{2,0}^{(Ca^2)} = \frac{\left\{\begin{array}{l}\left(a_0^{(20)}\lambda^3 + a_1^{(20)}\lambda^2 + a_2^{(20)}\lambda + a_3^{(20)}\right)\beta^3 + \left(b_0^{(20)}\lambda^3 + b_1^{(20)}\lambda^2 + b_2^{(20)}\lambda + b_3^{(20)}\right)\beta^2 \\ + \left(c_0^{(20)}\lambda^3 + c_1^{(20)}\lambda^2 + c_2^{(20)}\lambda + c_3^{(20)}\right)\beta - 25(19\lambda + 16)\left(601\lambda^2 + 893\lambda + 256\right)\end{array}\right\}}{1344(-5\beta - 5\beta\lambda + \beta k + 5 + 5\lambda)^3},$$

$$L_{2,2}^{(Ca^2)} = \frac{\left\{\begin{array}{l}\left(b_0^{(22)}\lambda^3 + b_1^{(22)}\lambda^2 + b_2^{(22)}\lambda + b_3^{(22)}\right)\beta^2 \\ + \left(c_0^{(22)}\lambda^3 + c_1^{(22)}\lambda^2 + c_2^{(22)}\lambda + c_3^{(22)}\right)\beta + 15(2\lambda + 3)(16 + 19\lambda)^2\end{array}\right\}}{1152(-5\beta - 5\beta\lambda + \beta k + 5 + 5\lambda)^2},$$

$$L_{4,0}^{(Ca^2)} = \frac{\left\{\begin{array}{l}\left(a_0^{(40)}\lambda^3 + a_1^{(40)}\lambda^2 + a_2^{(40)}\lambda + a_3^{(40)}\right)\beta^3 + \left(b_0^{(40)}\lambda^3 + b_1^{(40)}\lambda^2 + b_2^{(40)}\lambda + b_3^{(40)}\right)\beta^2 \\ + \left(c_0^{(40)}\lambda^3 + c_1^{(40)}\lambda^2 + c_2^{(40)}\lambda + c_3^{(40)}\right)\beta + 5(751\lambda + 656)(19\lambda + 16)(\lambda + 1)\end{array}\right\}}{1344(-5\beta - 5\beta\lambda + \beta k + 5 + 5\lambda)^2(\beta k - 9\beta - 9\beta\lambda + 9\lambda + 9)},$$

$$L_{4,4}^{(Ca^2)} = \frac{\left\{\begin{array}{l}\left(a_0^{(44)}\lambda^3 + a_1^{(44)}\lambda^2 + a_2^{(44)}\lambda + a_3^{(44)}\right)\beta^3 + \left(b_0^{(44)}\lambda^3 + b_1^{(44)}\lambda^2 + b_2^{(44)}\lambda + b_3^{(44)}\right)\beta^2 \\ + \left(c_0^{(44)}\lambda^3 + c_1^{(44)}\lambda^2 + c_2^{(44)}\lambda + c_3^{(44)}\right)\beta - 5(751\lambda + 656)(19\lambda + 16)(\lambda + 1)\end{array}\right\}}{32256(-5\beta - 5\beta\lambda + \beta k + 5 + 5\lambda)^2(\beta k - 9\beta - 9\beta\lambda + 9\lambda + 9)} \quad \text{(B7)}$$

where the constants in the above Eqs. (C1) and (C2) are given below

$$\left.\begin{array}{l}a_0^{(20)} = 285475, \; a_1^{(20)} = (664575 - 174100k), \\ a_2^{(20)} = (5300k^2 + 478800 - 270800k), \; a_3^{(20)} = 27200k^2 - 2400k^3 - 97600k + 102400, \\ b_0^{(20)} = -856425, \; b_1^{(20)} = (-1993725 + 348200k), \\ b_2^{(20)} = (541600k - 1436400 - 35200k^2), \; b_3^{(20)} = -26800k^2 - 307200 + 195200k, \\ c_0^{(20)} = 856425, \; c_1^{(20)} = (-174100k + 1993725), \\ c_2^{(20)} = (1436400 - 100k^2 - 270800k), \; c_3^{(20)} = -400k^2 + 307200 - 97600k,\end{array}\right\} \quad \text{(B8)}$$

$$\left.\begin{array}{l}b_0^{(22)} = 10830, \; b_1^{(22)} = (-4465k + 34485), \\ b_2^{(22)} = (35040 + 440k^2 - 10220k), \; b_3^{(22)} = 560k^2 + 11520 - 5440k, \\ c_0^{(22)} = -856425, \; c_1^{(22)} = (174100k - 1993725), \\ c_2^{(22)} = (-1436400 + 100k^2 + 270800k), \; c_3^{(22)} = -307200 + 97600k + 400k^2,\end{array}\right\} \quad \text{(B9)}$$

$$\left.\begin{array}{l} a_0^{(40)} = -71345, \ a_1^{(40)} = (-193745 + 38200k), \\ a_2^{(40)} = (-174880 + 69700k - 6570k^2), \ a_3^{(40)} = 31680k - 52480 + 360k^3 - 6040k, \\ b_0^{(40)} = 214035, \ b_1^{(40)} = (-76400k + 581235), \\ b_2^{(40)} = (6550k^2 - 139400k + 524640), \ b_3^{(40)} = -63360k + 6000k^2 + 157440, \\ c_0^{(40)} = -214035, \ c_1^{(40)} = (38200k - 581235), \\ c_2^{(40)} = (69700k - 524640 + 20k^2), \ c_3^{(40)} = -157440 + 31680k + 40k^2, \end{array}\right\} \quad (B10)$$

and

$$\left.\begin{array}{l} a_0^{(44)} = 71345, \ a_1^{(44)} = (193745 - 38200k), \\ a_2^{(44)} = (6570k^2 + 174880 - 69700k), \ a_3^{(44)} = -31680k - 360k^3 + 6040k^2 + 52480, \\ b_0^{(44)} = -214035, \ b_1^{(44)} = (76400k - 581235), \\ b_2^{(44)} = (-6550k^2 + 139400k - 524640), \ b_3^{(44)} = -157440 + 63360k - 6000k^2, \\ c_0^{(44)} = 214035, \ c_1^{(44)} = (581235 - 38200k), \\ c_2^{(44)} = (524640 - 69700k - 20k^2), \ c_3^{(44)} = 157440 - 31680k - 40k^2, \end{array}\right\} \quad (B11)$$

The velocity fields in both the phases is written as

$$\mathbf{u}_{i,0} = \begin{bmatrix} \dfrac{15}{4} \left\{ \dfrac{r(1-\beta)(-1+r^2)}{-5\beta - 5\lambda\beta + 5 + 5\lambda + k\beta} \right\} \sin(2\phi)\sin(\theta)^2 \mathbf{e}_r \\ + \dfrac{5}{8} \left\{ \dfrac{r(1-\beta)(-3+5r^2)}{-5\beta - 5\lambda\beta + 5 + 5\lambda + k\beta} \right\} \sin(2\phi)\sin(2\theta) \mathbf{e}_\theta \\ -\dfrac{1}{4} \dfrac{r \left\{ \begin{array}{l} -10\beta - 10\lambda\beta + 10 + 10\lambda + 2k\beta \\ -15\cos(2\phi)\beta + 15\cos(2\phi) + 25r^2\cos(2\phi)\beta - 25r^2\cos(2\phi) \end{array} \right\}}{-5\beta - 5\lambda\beta + 5 + 5\lambda + k\beta} \sin(\theta)\mathbf{e}_\varphi \end{bmatrix}, \quad (B12)$$

$$\mathbf{u}_{e,0} = \begin{bmatrix} \left\{ -\dfrac{5}{4r^4} \dfrac{\left[(-5\lambda+k-2)r^2 + 3\lambda - \dfrac{3}{5}k\right]\beta + (5\lambda+2)r^2 - 3\lambda}{(-5-5\lambda+k)\beta + 5 + 5\lambda} \sin(2\phi)\sin(\theta)^2 \right. \\ \left. + \dfrac{r}{2}\sin(\theta)^2 \sin(2\phi) \right\} \mathbf{e}_r \\ -\left( \dfrac{1}{4r^4} \dfrac{(-5\lambda+k)\beta + 5\lambda}{(-5-5\lambda+k)\beta + 5 + 5\lambda} \sin(2\theta)\sin(2\phi) + \dfrac{r}{4}\sin(2\theta)\sin(2\phi) \right) \mathbf{e}_\theta \\ -\left( \dfrac{1}{2r^4} \dfrac{(-5\lambda+k)\beta + 5\lambda}{(-5-5\lambda+k)\beta + 5 + 5\lambda} \sin(\theta)\cos(2\phi) - r\sin(\theta)\sin(\phi)^2 \right) \mathbf{e}_\varphi \end{bmatrix}$$ (B13)

The corresponding pressure field is given below

$$\left.\begin{aligned} p_{i,0} &= \dfrac{105 r^2 \lambda (1-\beta)}{(-20-20\lambda+4k)\beta + 20 + 20\lambda} \sin(2\phi)\sin(\theta)^2, \\ p_{e,0} &= -\dfrac{5}{2r^3} \left\{ \dfrac{(-5\lambda+k-2)\beta + 5\lambda + 2}{(-5-5\lambda+k)\beta + 5 + 5\lambda} \right\} \sin(2\phi)\sin(\theta)^2, \end{aligned}\right\}$$ (B14)

## Appendix C: Expression of the constants present in Eq. (45) and Eq. (47)

The constants present in the expression of effective extensional viscosity in Eq. (45), are given by

$$\left.\begin{aligned} & m_0^{(1)} = -475,\ m_1^{(1)} = (-1179 + 290k), \\ & m_2^{(1)} = (-732 - 59k^2 + 476k),\ m_3^{(1)} = (-64 + 144k + 4k^3 - 44k^2), \\ & m_0^{(2)} = -214035,\ m_1^{(2)} = (76400k - 581235), \\ & m_2^{(2)} = (-6550k^2 + 139400k - 524640),\ m_3^{(2)} = -157440 + 63360k - 6000k^2, \\ & m_0^{(3)} = 214035,\ m_1^{(3)} = (581235 - 38200k), \\ & m_2^{(3)} = (524640 - 69700k - 20k^2),\ m_3^{(3)} = 157440 - 31680k - 40k^2, \end{aligned}\right\}$$ (C1)

The constants present in the expression of $N_1$ and $N_2$ in Eq. (47) are given below

$$n_{1,0}^{(1)} = 361, \ n_{1,1}^{(1)} = (-152k + 608), \ n_{1,2}^{(1)} = (256 + 12k^2 - 128k)$$
$$n_{2,0}^{(1)} = -722, \ n_{2,1}^{(1)} = (152k - 1216), \ n_{2,2}^{(1)} = -512 + 4k^2 + 128k \quad (C2)$$

$$n_{1,0}^{(2)} = -5510, \ n_{1,1}^{(2)} = (-16230 + 3497k),$$
$$n_{1,2}^{(2)} = (-599k^2 - 19260 + 6916k), \ n_{1,3}^{(2)} = -656k^2 - 8000 + 4112k + 24k^3,$$
$$n_{2,0}^{(2)} = 16530, \ n_{2,1}^{(2)} = (48690 - 6994k),$$
$$n_{2,2}^{(2)} = (57780 - 13832k + 459k^2), \ n_{2,3}^{(2)} = -8224k + 576k^2 + 24000 + 28k^3, \quad (C3)$$
$$n_{3,0}^{(2)} = -16530, \ n_{3,1}^{(2)} = (-48690 + 3497k),$$
$$n_{3,2}^{(2)} = (-57780 + 140k^2 + 6916k), \ n_{3,3}^{(2)} = -24000 + 4112k + 80k^2,$$